\documentclass[%
 reprint,
 amsmath,amssymb,
 aps,
]{revtex4-2}

\usepackage{graphicx}%
\usepackage{dcolumn}%
\usepackage{booktabs}%
\usepackage{bm}%
\usepackage{hyperref}%
\usepackage{siunitx}%
\usepackage{physics}
\usepackage{xfrac}
\usepackage{csquotes}
\usepackage{listings}
\usepackage{float}

\DeclareMathOperator{\cov}{cov}
\DeclareMathOperator{\Var}{Var}
\DeclareMathOperator{\sign}{sign}
\DeclareMathOperator*{\argmin}{argmin}
\DeclareMathOperator*{\argmax}{argmax}
\DeclareMathOperator{\fitted}{fitted}
\DeclareMathOperator{\naive}{naive}
\DeclareMathOperator{\inflated}{inflated}
\DeclareMathOperator{\fmax}{f_{\max}}
\DeclareMathOperator{\fmaxopt}{optimal-f_{\max}}
\DeclareMathOperator{\f}{f}
\DeclareMathOperator{\pmin}{p_{\min}}

\begin{document}

\newcommand{\normal}[0]{\mathcal{N}}

\newcommand{\pt}{p_\textrm{T}}
\newcommand{\plong}{p_\textrm{L}}
\newcommand{\dalphat}{\delta\alpha_\textrm{T}}
\newcommand{\dphit}{\delta\phi_\textrm{T}}
\newcommand{\dpt}{\delta \pt}
\newcommand{\dplong}{\delta \plong}

\newcommand{\ibar}{\bar{\imath}}

\preprint{TBD}

\title{Hypothesis tests and model parameter estimation on data sets with missing correlation information}

\author{Lukas Koch}
 \email{lukas.koch@uni-mainz.de}
\affiliation{%
 Johannes Gutenberg University Mainz\\
 Institute of Physics - ETAP\\
 Staudingerweg 7\\
 55128 Mainz
}%

\date{\today}%

\begin{abstract}
Ideally, all analyses of normally distributed data should include the full covariance information between all data points.
In practice, the full covariance matrix between all data points is not always available.
Either because a result was published without a covariance matrix, or because one tries to combine multiple results from separate publications.
For simple hypothesis tests, it is possible to define robust test statistics that will behave conservatively in the presence on unknown correlations.
For model parameter fits, one can inflate the variance by a factor to ensure that things remain conservative at least up to a chosen confidence level.
This paper describes a class of robust test statistics for simple hypothesis tests, as well as an algorithm to determine the necessary inflation factor for model parameter fits and Goodness of Fit tests and composite hypothesis tests.
It then presents some example applications of the methods to real neutrino interaction data and model comparisons.
\end{abstract}

\maketitle

\section{Introduction}

The ``standard'' way of releasing measurements is often in the form of a central value $\bm{x}$ plus an uncertainty encoded in a covariance matrix $S$.
This constitutes an implicit approximation of the Likelihood function (or in the case of Bayesian statistics the posterior probability) as a multivariate normal distribution.
It is important to use the full covariance matrix when using the results for any statistical analysis, since ignoring it can lead to very wrong conclusions, e.g. when determining whether or not a specific prediction is compatible with the data.

Unfortunately, the full covariance matrix is not always available.
\cite{Koch2021} discusses how to do simple model tests with data sets that have no information about the correlation at all, by choosing alternative test statistics that are robust against the presence of correlations.
This is the case, e.g., when dealing with published results that do not include a covariance matrix.

In \autoref{sec:fitted} and following we will discuss how to generalise one such test statistic to the case of known blocks of covariance, but unknown correlations between these blocks.
This is the case, e.g., when doing a statistical test of a model against multiple published results, each of which with a full covariance matrix, but no information about the correlations between the results.

The alternative test statistics are suitable for simple hypothesis tests, but less so for doing model parameter fits to the data.
For this we will introduce a derating factor for the covariance in \autoref{sec:derate} and following.
This allows the ``usual'' procedures to be used while ensuring that the result will remain conservative up to a chosen confidence level.

This approach is conceptually similar to the flat out doubling of variance as discussed in \cite{Meng2022}, and the $S$-factor employed by the Particle Data Group\cite[Sec.\,5.2]{Navas2024}.
The former is intended to cover any shortcoming in the statistical treatment, by erring on the side of making weaker statements.
The latter inflates the uncertainties based on the actual Goodness of Fit (GoF) $\chi^2$ score, and forces it to be at most $1/(N-1)$, with $N$ the number of combined measurements.
It also still assumes that the combined measurements are statistically independent.
Since the GoF is a random variable, this means that the derating factor is also random.
The method presented here, on the other hand, bases the derating factor on the worst-case scenario of potential correlations between data sets.
It only indirectly depends on the actual data, as the best-fit points determine the details of the linear approximation of the fitted models.

Finally, in \autoref{sec:GoF} we will discuss how to use the derating method not just to inflate parameter uncertainties, but also to do Goodness of Fit and composite hypothesis tests.

\section{Generalisation of the fitted test statistic}
\label{sec:fitted}

The \enquote{fitted} test statistic described in \cite{Koch2021} treats the unknown covariance elements as nuisance parameters and minimizes the Mahalanobis distance\cite{Mahalanobis1936} (M-distance) over the possible covariance space.
It turns out that this minimum possible M-distance is equivalent to the largest among the single-bin z-scores (data to model discrepancy in ``sigmas'').
This can be generalised to scenarios where blocks of covariances are known, but the correlations between those blocks are unknown:
\begin{align}
    S = \mqty(S_{11} & S_{12} & S_{13} &  \\
              S^T_{12} & S_{22} & S_{23} & \cdots \\
              S^T_{13} & S^T_{23} & S_{33} &  \\
                       & \vdots   &        & \ddots{} ) \text{,}
\end{align}
with the $N$ known square matrices $S_{ii}$ of sizes $N_i \cross N_i$ and the unknown off-diagonal submatrices $S_{ij}$ of sizes $N_i \cross N_j$.
The scenario investigated in \cite{Koch2021} is a special case, where all known matrices are of size $1 \cross 1$.

For simplicity of notation later on, we will assume that the blocks of known covariance are ordered in ascending blockwise M-distance, $D_i$.
\begin{align}
    D^2_i &= (\bm{x}_i - \bm\mu_i)^T S_{ii}^{-1} (\bm{x}_i - \bm\mu_i) , \\
    D^2_{i+1} &> D^2_i \qfor i \in \{ 1, 2, \dots, N-1 \}. \label{eq:order}
\end{align}

The \enquote{fitted} test statistic is now the minimum possible squared Mahalanobis distance,
minimized over all possible off-diagonal matrices:
\begin{align}
    \fitted(\bm{x} | \bm\mu, \bm{S}) &= \min_{S} (\bm{x} - \bm\mu)^T S^{-1} (\bm{x} - \bm\mu) \\
      &= \max_i (\bm{x_i} - \bm\mu_i)^T S_{ii}^{-1} (\bm{x_i} - \bm\mu_i) \text{,} \label{eq:minmax}
      = D^2_N
\end{align}
where $\bm{S}$ is the vector of known square matrices, and the minimisation over $S$ only varies the unknown off-diagonal matrices.
It is equal to choosing the maximum among the blockwise M-distances, where $\bm{x}_i$, and $\bm\mu_i$ are the vectors of data points and expectation values corresponding to the known covariance matrix $S_{ii}$.

The proof of \autoref{eq:minmax} is analogous to the proof of the special case in \cite{Koch2021}.
The M-distance is conserved when doing linear coordinate transforms.
We can transform the variable space so that the last block of variables $\bm{x}_{N}$ is uncorrelated to any other variables:
\begin{widetext}
\begin{align}
    \bm{y} = A \bm{x} &= \mqty(I_{N_1} & 0 & & -S_{1N} S^{-1}_{NN} \\
                    0 & I_{N_2} & \cdots & -S_{2N} S^{-1}_{NN} \\
                      & \vdots & & \vdots \\
                    0 & 0 & \cdots & I_{N_N}
              ) \bm{x} , \\
    \cov(\bm{y}) = S^y &= ASA^T = \mqty(S^y_{11} & S^y_{12} & & 0 \\
                                S^{yT}_{12} & S^y_{22} & \cdots & 0 \\
                                 & \vdots & & \vdots \\
                                0 & 0 & \cdots & S_{NN}
              ) , \\
    \expval{\bm{y}} = \bm\mu^y &= A \bm\mu , \\
    \expval{\bm{y}_i} = \bm\mu^y_i &= \bm\mu_i - S_{iN}S_{NN}^{-1}\bm\mu_N \qfor i \in \{ 1, 2, \dots, N-1 \}, \\
    \expval{\bm{y}_N} = \bm\mu^y_N &= \bm\mu_N ,
\end{align}
\end{widetext}
with the new blockwise covariances $S^y_{ij}$ for $i,j \in \{ 1, 2, \dots, N-1 \}$,
whose actual values do not matter for this proof.
The contribution of $\bm{y}_N$ to the total M-distance in this coordinate system is constant,
because $S_{NN}$ is known and $\bm{y}_N$ is not correlated with any of the other variables.

The contribution of the other $\bm{y}_i$ can be eliminated by tuning the $S_{iN}$ such that the expectation values are identical to the actually measured values:
\begin{align}
    \bm\mu^y_i = \bm\mu_i - S_{iN}S_{NN}^{-1}\bm\mu_N &\overset{!}{=} \bm{x}_i - S_{iN}S_{NN}^{-1}\bm{x}_N = \bm{y_i} , \\
    (\bm{x}_i - \bm\mu_i) &= S_{iN} S_{NN}^{-1} (\bm{x}_N - \bm\mu_N),
\end{align}
which can be achieved by setting
\begin{widetext}
\begin{align}
    S_{iN} &= \frac{(\bm{x}_i - \bm\mu_i) (\bm{x}_N - \bm\mu_N)^T}
                {(\bm{x}_N - \bm\mu_N)^T S_{NN}^{-1} (\bm{x}_N - \bm\mu_N)}
            = \frac{(\bm{x}_i - \bm\mu_i) (\bm{x}_N - \bm\mu_N)^T}
                {D^2_N}. \label{eq:mincov}
\end{align}
\end{widetext}

To show that the resulting covariance matrix $S$ is positive definite,
we can use Sylvester's criterion\cite{Gilbert1991}, which states that a matrix is positive definite
if and only if all leading principal minors of the matrix are positive.
This can be done by computing the determinant using the properties of partitioned matrices and the matrix determinant lemma\cite[Sec.\,13.3\,\&\,18.1]{Harville1997}.
Let $A_{[k,l]}$ denote the submatrix of $A$ with the last $k$ rows and $l$ columns removed.
We then have
\begin{widetext}
\begin{align}
    \det(S_{[k,k]}) =& \mqty|S_{\dagger\dagger} & S_{\dagger N[0,k]} \\ (S_{\dagger N[0,k]})^T & S_{NN[k,k]}| = \det(S_{NN[k,k]}) \det(S_{\dagger\dagger} - S_{\dagger N[0,k]}(S_{NN[k,k]})^{-1}(S_{\dagger N[0,k]})^T) \nonumber\\
        =& \det(S_{NN[k,k]}) \nonumber\\
         & \det(
            S_{\dagger\dagger}
            - \frac{(\bm{x}_\dagger - \bm\mu_\dagger)(\bm{x}_{N} - \bm\mu_{N})^T_{[0,k]}}{D^2_{N}}
              (S_{NN[k,k]})^{-1}
              \frac{(\bm{x}_{N} - \bm\mu_{N})_{[k,0]}(\bm{x}_\dagger - \bm\mu_\dagger)^T}{D^2_{N}}
            ) \nonumber\\
        =& \det(S_{NN[k,k]}) \det(
            S_{\dagger\dagger}
            - \frac{(\bm{x}_\dagger - \bm\mu_\dagger)(\bm{x}_\dagger - \bm\mu_i)^T}{\gamma D^2_N} ) \nonumber\\
        =& \det(S_{NN[k,k]}) \det(S_{\dagger\dagger}) \qty( 1 - \frac{(\bm{x}_\dagger - \bm\mu_\dagger)^T S_{\dagger\dagger}^{-1}(\bm{x}_\dagger - \bm\mu_\dagger)}{\gamma D^2_N} ) \nonumber\\
        =& \det(S_{NN[k,k]}) \det(S_{\dagger\dagger}) \qty(1 - \frac{D^2_\dagger}{\gamma D^2_N}).
\end{align}
\end{widetext}
Here $\bm{x}_\dagger$ is a vector of all the data points except the $N$th block, and $\gamma = D^2_N / ( (\bm{x}_{N} - \bm\mu_{N})_{[0,k]}^T (S_{NN[k,k]})^{-1}(\bm{x}_{N} - \bm\mu_{N})_{[k,0]}) \ge 1$.
The determinant of $S_{NN[k,k]}$ is positive by definition for all non-negative $k < N_N$.
So, if $S_{\dagger\dagger}$ is positive definite and the blockwise M-distance $D^2_N$ is larger than the M-distance of the remaining data without the $N$th block $D^2_\dagger$,
the determinant of $S_{[k,k]}$ will be positive for all $k < N_N$ and the choice of $S_{iN}$ in \autoref{eq:mincov} is valid.
Since the off-diagonal blocks in $S_{\dagger\dagger}$ are still undetermined,
these two conditions can be proven by applying the above steps recursively to $S_{\dagger\dagger}$, setting $D^2_\dagger$ to be the largest among the remaining $D^2_i < D^2_N$, and eliminating the block with the largest block M-distance at each step. This is repeated until only $S_{11}$ remains, which is positive definite by definition.
In the infinitely unlikely case that two block M-distances are exactly the same, the resulting covariance matrix S will be positive semi-definite instead.

Since the test statistic is reduced to the maximum of a fixed number of random variables,
the expected distribution assuming no correlations between those variables can be easily calculated.
The overall Cumulative Distribution Function (CDF) is simply the product of the independent CDFs.
Continuing the tradition established in \cite{Koch2021}, let us call this distribution the Cee-squared distribution.
A basic implementation in Python is available as part of the Python package NuStatTools, which can be downloaded from Zenodo, GitHub, and PyPI\cite{Koch2024}.

A general argument about the robustness of this type of test statistic is made in \autoref{sec:fmax}.
To illustrate the performance,
we will use 10-dimensional multivariate Gaussian toy data with an expectation value of 0,
and a covariance
\begin{equation}
    V = \mqty*(I_5 & \rho I_5 \\
               \rho I_5 & I_5). \label{eq:toy-cov}
\end{equation}
I.e. there are two uncorrelated blocks of 5 variables each, and a one-to-one correlation between
variable pairs of those blocks. The ``correlation factor'' $\rho$ is set to $0$, $0.5$, $0.9$, and $0.99$ in the different data sets. We use the letter $V$ to distinguish the real covariance from the assumed covariance $S$.

\begin{figure*}
    \centering
    \includegraphics[width=0.49\textwidth]{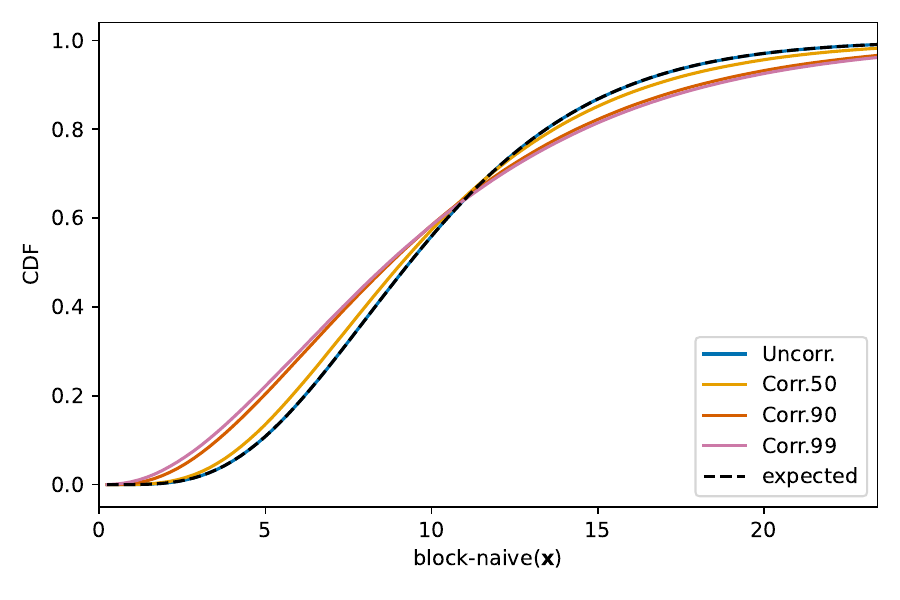}
    \includegraphics[width=0.49\textwidth]{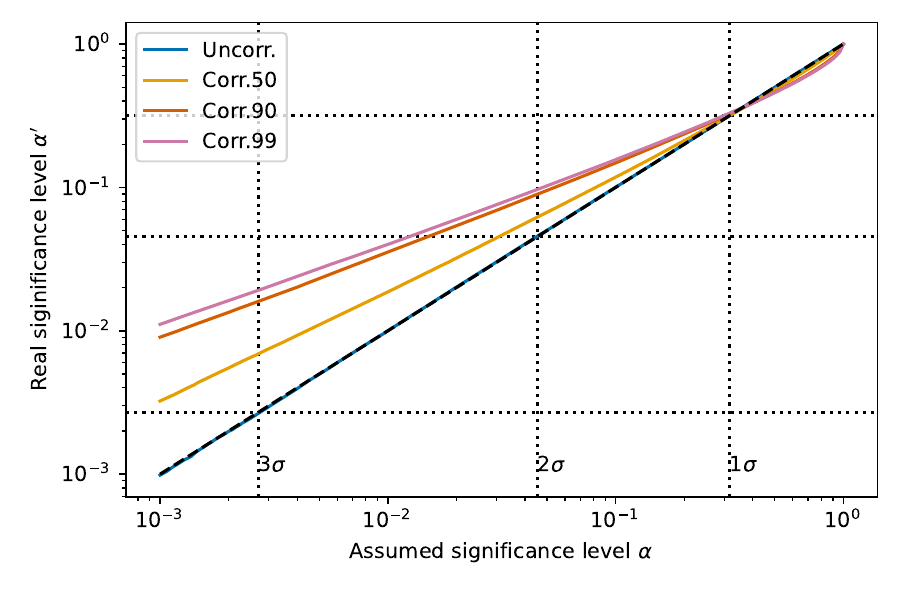}
    \caption{\label{fig:block-naive}%
        CDFs (left) for the ``naive'' squared M-distance test statistic for different levels of correlations in the data.
        When using the uncorrelated CDF to calculate the assumed significance level (or p-value) of a value of the statistic, the actual level will differ from the assumption depending on the correlations (right).
        Where the real significance level is larger than the assumed significance level (the real significance is weaker than the assumed one), the test statistic shows undercoverage.
        This corresponds to the region in the CDF where the real CDF for a given value is lower than the expected one.
    }
\end{figure*}

\begin{figure*}
    \centering
    \includegraphics[width=0.49\textwidth]{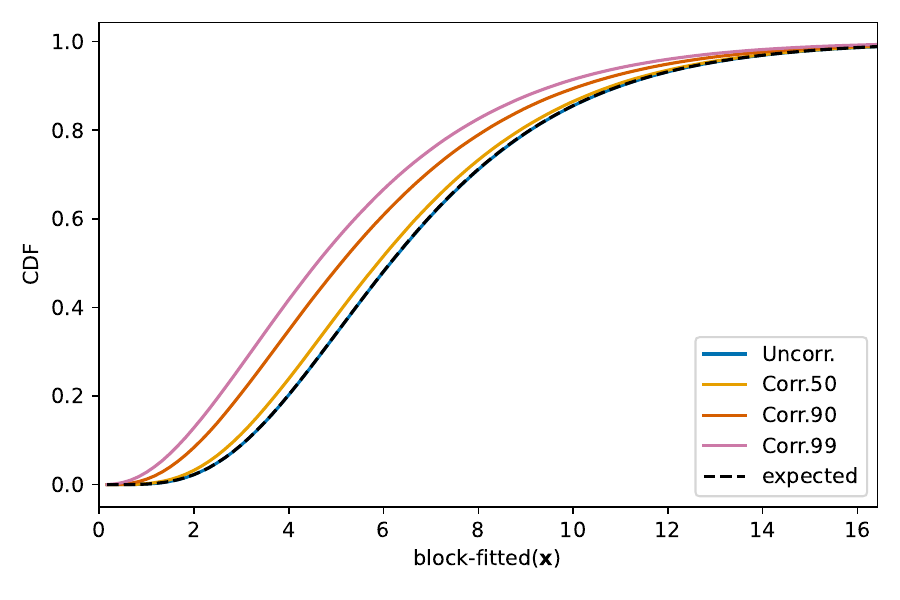}
    \includegraphics[width=0.49\textwidth]{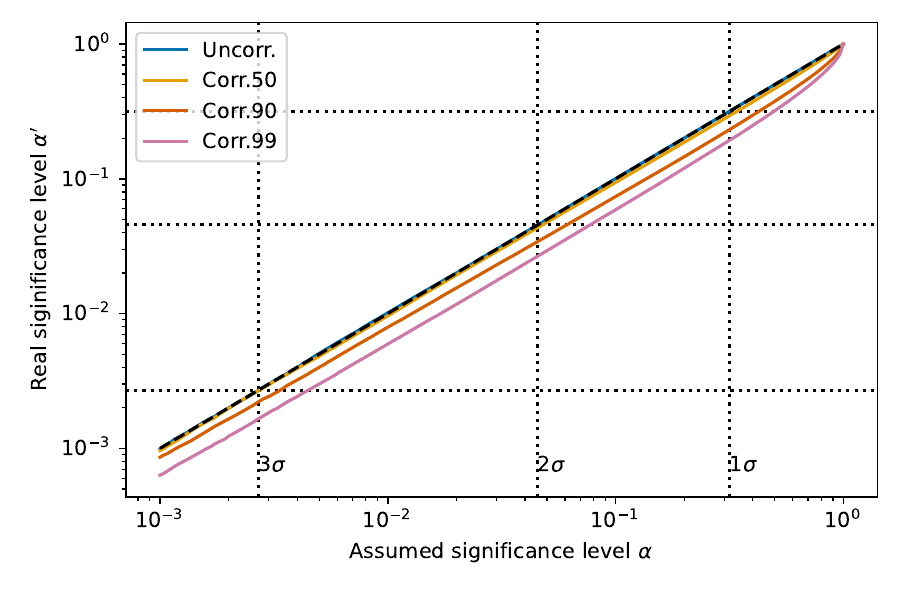}
    \caption{\label{fig:block-fitted}%
        CDFs (left) for the ``fitted'' test statistic for different levels of correlations in the data.
        When using the uncorrelated CDF to calculate the assumed significance level (or p-value) of a value of the statistic, the actual level will differ from the assumption depending on the correlations (right).
        In the presence of correlations, the real significance is consistently higher (the significance level is lower) than the assumption. This means the uncertainties are overestimated and the statistic behaves conservatively.
    }
\end{figure*}

\autoref{fig:block-naive} shows how the ``naive'' application of the M-distance without taking into account any correlations performs. In agreement with what was shown in \cite{Koch2021},
the real significance in the presence of unaccounted correlations can be much weaker than the assumed significance if one expects a chi-squared distributed test statistic.
We are seeing undercoverage of the confidence regions for anything above about the one-sigma level.
In contrast, \autoref{fig:block-fitted} shows how the $\fitted$ test statistic performs on the same data.
It is consistently conservative for all significance levels at all levels of correlation.

\section{Application to neutrino model tunes}

Since the $\fitted$ test statistic is calculated by taking the maximum squared M-distance among the combined measurements, it can be very easily applied to combine existing tests of the same model. The authors of \cite{Filali2024} compare a selection of neutrino interaction models against multiple cross-section measurements. Since the correlations between the different experimental results are unknown, they take care not to make any quantitative claims about the overall fit of the models to the combination of different data sets. With the robust test statistic we can make such statements.

\begin{table*}
    \centering
    \caption{Results of model tests against single data sets, as described in \cite{Filali2024}.
    P-values in parenthesis are taken from Table~III and Table~V.
    Squared M-distances are taken from the respective plots.
    }
    \label{tab:filali}
    \resizebox{\textwidth}{!}{
    \begin{tabular}{lccc ccc cc cc}
\toprule
Measurement                           & $N_{bins}$ & SF/SF* & LFG & RFG & More 2p2h & More FSI & Less FSI & More $\pi$ abs. & Less $\pi$ abs. & GENIE\\ 
\midrule
T2K $\dalphat$                        & 8          & 19.59 (0.01) & (0.00) & (0.00) & (0.00)  & 22.33 (0.00) & 18.79 (0.02) & (0.06)  & (0.02)  \\
$\dpt$                            & 8          & 13.91 (0.08) & 5.61 (0.69) & 85.55 (0.00) & 34.36 (0.00) & 17.89 (0.02) & 14.68 (0.07) & (0.00)  & (0.18) & 7.62 (0.47) \\
\addlinespace
MINERvA $\dalphat$                    & 12         & (0.00) & (0.00) & (0.00) & (0.00)  & (0.00) & (0.06) & (0.00)  & (0.00)  \\
$\dpt$                        & 24         & 64.26 (0.00) & 71.10 (0.00) & 415.29 (0.00) & 92.50 (0.00)  & 58.98 (0.00) & 96.25 (0.00) & 109.36 (0.00)  & 63.65 (0.00) & 49.15 (0.00) \\
$p_N$                         & 24         & 93.17 (0.00) & 101.33 (0.00) & 549.46 (0.00) & (0.00) & (0.00) & (0.00) & (0.00)  & (0.00)  \\
\addlinespace
MicroBooNE $\dalphat$                 & 7          & 17.03 (0.02) & 6.78 (0.45) & 5.33 (0.62) & (0.07)  & 10.10 (0.18) & 25.42 (0.00) & (0.02)  & (0.01)  \\
$\dpt$                     & 13         & 19.69 (0.10) & 13.42 (0.42) & 31.52 (0.00) & 14.67 (0.33) & 16.30 (0.23) & 24.75 (0.02) & (0.13)  & (0.10) & 9.59 (0.73) \\
$\dpt$ low $\dalphat$      & 11         & 13.51 (0.26) & 14.00 (0.23) & 16.06 (0.14) & 11.93 (0.37) & 11.08 (0.44) & 17.45 (0.10) & (0.28)  & (0.24) & 10.52 (0.28) \\
$\dpt$ mid-low $\dalphat$  & 12         & 19.63 (0.07) & 12.53 (0.40) & 16.06 (0.19) & 15.17 (0.23) & 12.85 (0.38) & 28.54 (0.00) & (0.08)  & (0.06) & 9.81 (0.41) \\
$\dpt$ mid-high $\dalphat$ & 13         & 22.81 (0.04) & 16.29 (0.23) & 25.32 (0.02) & 18.02 (0.16) & 16.54 (0.22) & 29.62 (0.01) & (0.05)  & (0.04) & 14.84 (0.32) \\
$\dpt$ high $\dalphat$     & 13         & 23.62 (0.03) & 18.68 (0.13) & 20.78 (0.08) & 19.02 (0.12) & 20.42 (0.09) & 28.22 (0.01) & (0.04)  & (0.03) & 17.43 (0.18) \\
\bottomrule
    \end{tabular}}
\end{table*}

\autoref{tab:filali} shows the results for model comparisons against single data sets as reported in the paper.
To make a quantitative statement about a model's tensions with multiple data sets, all we need to do is take the largest among the relevant squared M-distances and calculate its significance using the correct Cee-squared distribution.
For example, the Spectral Function (SF) model is excluded by the T2K $\dalphat$ measurement, while it is compatible with the T2K $\dpt$ measurement.
The maximum squared M-distance is $19.59$, which yields a p-value of $0.02$ when compared with a Cee-squared distribution for two blocks of 8 variables.
So the combination of the two measurements excludes the model at the 98\% confidence level.

\begin{table*}
    \centering
    \caption{P values calculated with the $\fitted$ test statistic for various combinations of measurements from \autoref{tab:filali}.
    Only measurements with a squared M-distance are used.
    The ``MicroBooNE all'' group includes the four ``(mid-)low/high $\dalphat$'' measurements, while the ``MicroBooNE'' one does not.
    There is actually a full covariance matrix available between those four measurements, but it was not used in \cite{Filali2024}, nor here.
    The numbers in parentheses show how many measurements were combined.
    Groupings that contain only one valid measurement are omitted, as the p-value is identical to the one in \autoref{tab:filali}.}
    \label{tab:fitted}
    \begin{tabular*}{\textwidth}{@{\extracolsep{\fill}}lccccccc}
\toprule
Measurements & SF/SF* & LFG & RFG & More 2p2h & More FSI & Less FSI & GENIE \\
\midrule
T2K & 0.024 (2) &  &  &  & 0.009 (2) & 0.032 (2) &  \\
MINERvA & 0.000 (2) & 0.000 (2) & 0.000 (2) &  &  &  &  \\
MicroBooNE & 0.109 (2) & 0.452 (2) & 0.003 (2) &  & 0.251 (2) & 0.021 (2) &  \\
MicroBooNE all & 0.135 (6) & 0.456 (6) & 0.011 (6) & 0.421 (5) & 0.312 (6) & 0.021 (6) & 0.569 (5) \\
T2K+MicroBooNE & 0.129 (4) & 0.506 (3) & 0.000 (3) & 0.001 (2) & 0.061 (4) & 0.024 (4) & 0.808 (2) \\
T2K+MicroBooNE all & 0.140 (8) & 0.466 (7) & 0.000 (7) & 0.004 (6) & 0.200 (8) & 0.021 (8) & 0.580 (6) \\
all & 0.000 (10) & 0.000 (9) & 0.000 (9) & 0.000 (7) & 0.000 (9) & 0.000 (9) & 0.002 (7) \\
\bottomrule
    \end{tabular*}
\end{table*}

\autoref{tab:fitted} shows more results of different combinations of measurements for the tested models.
Many models are capable of describing the combined T2K and MicroBooNE data very well,
but in combination with the MINERvA results all models are excluded at the $99.7\%$ confidence level or stronger.

\section{Improving the statistical power of the fitted test statistic}
\label{sec:fmax}

One might wonder whether it makes sense to accept a model that is strongly excluded by one measurement, but accepted by another with more degrees of freedom.
E.g. if there is one measurement with $D_M^2 = 50$ and $df = 5$, combined with a second measurement with $D_M^2 = 45$ and $df=40$,
the $\fitted$ test statistic would be $50$, and the Cee-squared distribution would be very similar to a $\chi^2(40)$ distribution.
So, despite being very much excluded by the first measurement, the mere presence of the second measurement would mean that the model is accepted in the combined test.

However, this ``dilution'' of significance is not unique to the $\fitted$ test statistic.
We are dealing with the same kind of effect when doing the regular, uncorrelated combination of squared M-distances.
Here as well, a very significant result with few degrees of freedom can become irrelevant when combined with another result with many degrees of freedom.
In a sense, this is simply the ``look elsewhere effect'' in action.

\begin{figure}
    \centering
    \includegraphics[width=0.45\textwidth]{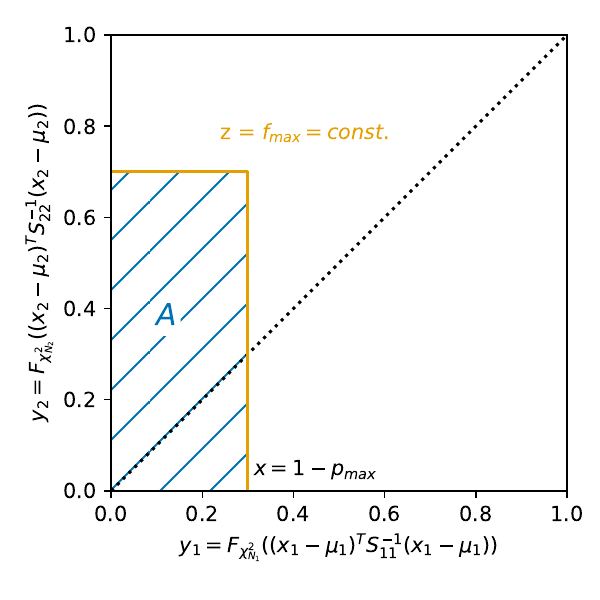}
    \caption{Illustration of the robustness of $\fmax$ test statistics for 2 blocks.
    If there are no correlations between the blocks, the $y$ variables will be independently uniformly distributed, and the expected CDF for the $\fmax$ statistic as a function of $z$ is equal to the area $A = \prod_i (1 - p_i)$.
    In the presence of unaccounted correlations, the ``worst case'' is if the $y$ variables are $100\%$ correlated, and all probability is concentrated along the diagonal line.
    In this case, the actual CDF is equal to $x = (1-p_{\max}) > A$.
    This means the assumed p-value $1 - A$ is bigger than the real p-value $p_{\max}$, and the $\fmax$ test statistic is conservative.
    }
    \label{fig:f-max-cdf}
\end{figure}

Still, it is possible to improve the statistical power of the test statistic.
We can generalize the fitted test statistic by investigating any test statistic that is the maximum of strictly increasing functions of the block M-distances:
\begin{align}
    \fmax(\bm{x}| \bm{\mu}, \bm{S}) = \max_i \f_i((\bm{x_i} - \bm\mu_i)^T S_{ii}^{-1} (\bm{x_i} - \bm\mu_i)).
\end{align}
Curves of constant $\fmax$ are rectangles in the CDF space of the squared block M-distances,
as the CDF is itself a strictly increasing function.
This means all test statistics of this kind should be robust against unknown correlations, as illustrated in \autoref{fig:f-max-cdf}.
Note that the functions $\f_i$ can be different for each block.
In fact, all $\fmax$ statistics with identical $\f_i$ are equivalent.
The $\fitted$ test statistic is such the special case where all $\f_i(x) = x$.
More generally, $\fmax$ statistics are equivalent, if the functions $\f'_i$ of one statistic can be expressed as the composition of a single strictly increasing function $g$ with the $\f_i$ of the other statistic:
\begin{align}
    \f'_i(x) = g(\f_i(x)).
\end{align}
The expected CDF of the $\fmax$ statistics depends on the functions $\f_i$:
\begin{align}
    F_{\fmax}(z) = \prod_i F_{\chi^2(N_i)}\qty( \f_i^{-1}(z) ),
\end{align}
where $\f_i^{-1}$ is the inverse of $\f_i$.

One possible alternative $\fmax$ test statistic would be not to select the largest M-distance, but the smallest p-value among the $N$ combined measurements, $p_{\min}$.
This is achieved by having $\f_i(x) = F_{\chi^2(N_i)}(x) = 1 - p_i$.
This way, the most significant measurement is always the relevant one, irrespective of the number of degrees of freedom.
The CDF calculation for this test statistic is very simple:
\begin{align}
    F_{\pmin}(q) &= (1-p_{\min})^N,
\end{align}
Note that the combined p-value $1-F_{\pmin}$ is still ``diluted'' by the number of combined measurements.
For small $p_{\min}$, it can be approximated as
\begin{align}
    p &= 1-F_{\pmin} \approx N p_{\min}.
\end{align}

\begin{table*}
    \centering
    \caption{P values calculated with the $\pmin$ test statistic for various combinations of measurements from \autoref{tab:filali}.
    Only measurements with a squared M-distance are used.
    The ``MicroBooNE all'' group includes the four ``(mid-)low/high $\dalphat$'' measurements, while the ``MicroBooNE'' one does not.
    There is actually a full covariance matrix available between those four measurements, but it was not used in \cite{Filali2024}, nor here.
    The numbers in parentheses show how many measurements were combined.
    Groupings that contain only one valid measurement are omitted, as the p-value is identical to the one in \autoref{tab:filali}.}
    \label{tab:p-min}
    \begin{tabular*}{\textwidth}{@{\extracolsep{\fill}}lccccccc}
\toprule
Measurements & SF/SF* & LFG & RFG & More 2p2h & More FSI & Less FSI & GENIE \\
\midrule
T2K & 0.024 (2) &  &  &  & 0.009 (2) & 0.032 (2) &  \\
MINERvA & 0.000 (2) & 0.000 (2) & 0.000 (2) &  &  &  &  \\
MicroBooNE & 0.034 (2) & 0.659 (2) & 0.006 (2) &  & 0.332 (2) & 0.001 (2) &  \\
MicroBooNE all & 0.099 (6) & 0.576 (6) & 0.017 (6) & 0.480 (5) & 0.414 (6) & 0.004 (6) & 0.630 (5) \\
T2K+MicroBooNE & 0.047 (4) & 0.801 (3) & 0.000 (3) & 0.000 (2) & 0.017 (4) & 0.003 (4) & 0.721 (2) \\
T2K+MicroBooNE all & 0.092 (8) & 0.633 (7) & 0.000 (7) & 0.000 (6) & 0.034 (8) & 0.005 (8) & 0.697 (6) \\
all & 0.000 (10) & 0.000 (9) & 0.000 (9) & 0.000 (7) & 0.001 (9) & 0.000 (9) & 0.013 (7) \\
\bottomrule
    \end{tabular*}
\end{table*}

\autoref{tab:p-min} shows the resulting p-values from combining the same measurements as in \autoref{tab:fitted} using the $\pmin$ test statistic.
As expected, it tends to produce stronger exclusions where the combination includes a measurement with significant exclusion, but small number of bins.
But it produces larger p-values where a measurement with large number of bins already provides the strongest block p-value.

In order to find a $\fmax$ statistic with a large statistical power, we can try to minimize the maximum M-distance that is still accepted at a given significance level, assuming no unaccounted correlations.
The motivation for this is that the likelihood ratio is the most powerful test statistic for simple hypothesis tests,
and the likelihood of a Gaussian distribution is a strictly decreasing function of the M-distance.
The $\naive$ test statistic would actually perform the best any test statistic could, judging by this criterion.
But as we have seen, it is unfortunately not robust against unknown correlations.

The maximum M-distance for a given p-value is always at the corner in the M-distance space where all $N$ block M-distances are at the maximum they can be for a given value of $\fmax$, and it is a strictly increasing function of the CDF of the test statistic.
So rather than maximizing the M-distance for a given p-value, we can also maximize the CDF as a function of the maximum M-distance at the corner point $D^2_i$ of the accepted phase-space:
\begin{widetext}
\begin{align}
    D^2 &= \sum_i D_i^2 \overset{!}{=} const.,\\
    F_{\fmax} &= \prod_i F_{\chi^2(N_i)}(D_i^2) ,\\
    \dv{F_{\fmax}}{D^2_i} &= f_{\chi^2(N_i)}(D_i^2) \prod_{j \ne i} F_{\chi^2(N_j)}(D_j^2),\\
    \dv{F_{\fmax}}{D^2_i} - \dv{F_{\fmax}}{D^2_j} &\overset{!}{=} 0 \quad\forall~i,j, \label{eq:crit}\\
    f_{\chi^2(N_i)}(D_i^2) F_{\chi^2(N_j)}(D_j^2) &= f_{\chi^2(N_j)}(D_j^2) F_{\chi^2(N_i)}(D_i^2) \quad\forall~i,j,\\
    \frac{F_{\chi^2(N_i)}(D_i^2)}{f_{\chi^2(N_i)}(D_i^2)} &= \frac{F_{\chi^2(N_j)}(D_j^2)}{f_{\chi^2(N_j)}(D_j^2)} \quad\forall~i,j ,\label{eq:optimal}
\end{align}
\end{widetext}
with $f_{\chi^2(N_i)}$ the PDF of the $\chi^2(N_i)$ distribution.
The criterion in \autoref{eq:crit} stems from the fact that we are looking for an extremum in the CDF of the $\fmax$ statistic, on the surface of constant $D^2$.
$D^2$ is guaranteed to be constant in the variation of $D^2_i$ if we vary any $D^2_j$ at the same time in the opposite direction.

\begin{figure}
    \centering
    \includegraphics[width=0.45\textwidth]{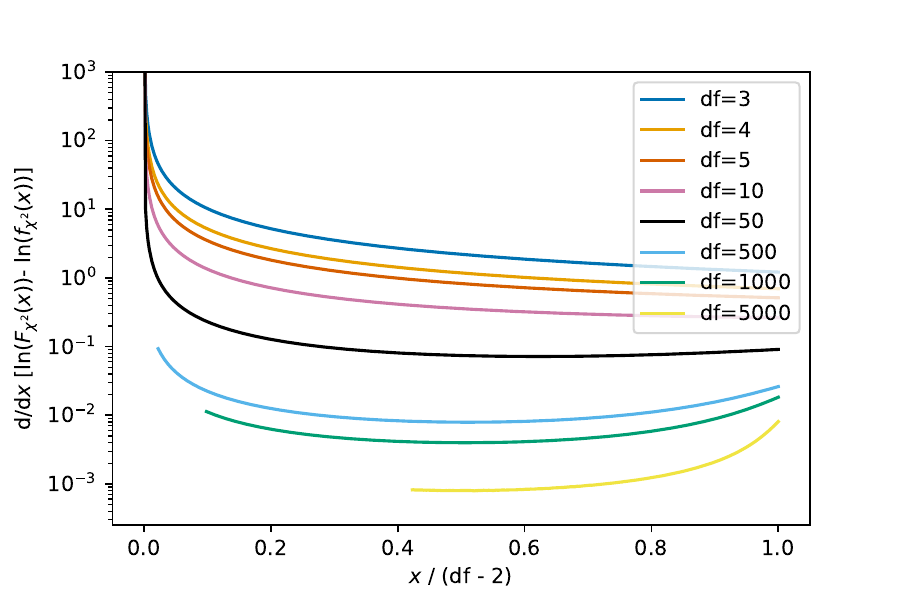}
    \caption{Derivative of the functions used for the $\fmaxopt$ test statistic for different number of degrees of freedom.
    Only the parameter range up to the mode of the $\chi^2$ distribution is shown.
    The derivative is strictly positive.
    For large number of degrees of freedom, limits in the numerical precision mean that the CDF is evaluated as exactly 0 for small $x$.
    This limits the range where the derivative can be calculated.
    The derivative is always positive, so the function is suitable for an $\fmax$ statistic.
    }
    \label{fig:positive}
\end{figure}

\autoref{eq:optimal} describes the condition for the maximum CDF at the corner of the acceptable phase space in terms of functions of $D^2_i$ that need to be identical in value.
This is exactly what the $\fmax$ statistic implements, so we could use those ratios of $\chi^2$ CDF and PDF directly as the $\f_i$, if it is strictly increasing.
We can see that it is definitely increasing as long as the PDF is decreasing, i.e. for $D^2_i$ values larger than the mode of the distribution.
For values below the mode, we can check that it is increasing numerically, as is shown in \autoref{fig:positive}.

The fact that the PDF gets arbitrarily small for high values of $D^2_i$ can lead to numerical issues in calculating this test statistic.
To get around this we apply a logarithm:
\begin{align}
    \f_i(D^2_i) = \ln(F_{\chi^2(N_i)}(D_i^2)) - \ln(f_{\chi^2(N_i)}(D_i^2)) .
\end{align}
Implementations of the logarithm of the $\chi^2$ PDF and CDF are readily available, e.g. as part of the Python package SciPy\cite{Virtanen2020}.
We call the resulting test statistic the $\fmaxopt$ statistic.
It is very similar to the $\pmin$ statistic, where $\f_i = F_{\chi^2(N_i)}$.
As such, the resulting p-values shown in \autoref{tab:optimal} are quite similar to those shown in \autoref{tab:p-min}.

\begin{table*}
    \centering
    \caption{P values calculated with the $\fmaxopt$ test statistic for various combinations of measurements from \autoref{tab:filali}.
    Only measurements with a squared M-distance are used.
    The ``MicroBooNE all'' group includes the four ``(mid-)low/high $\dalphat$'' measurements, while the ``MicroBooNE'' one does not.
    There is actually a full covariance matrix available between those four measurements, but it was not used in \cite{Filali2024}, nor here.
    The numbers in parentheses show how many measurements were combined.
    Groupings that contain only one valid measurement are omitted, as the p-value is identical to the one in \autoref{tab:filali}.}
    \label{tab:optimal}
    \begin{tabular*}{\linewidth}{@{\extracolsep{\fill}}lccccccc}
\toprule
Measurements & SF/SF* & LFG & RFG & More 2p2h & More FSI & Less FSI & GENIE \\
\midrule
T2K & 0.024 (2) &  &  &  & 0.009 (2) & 0.032 (2) &  \\
MINERvA & 0.000 (2) & 0.000 (2) & 0.000 (2) &  &  &  &  \\
MicroBooNE & 0.038 (2) & 0.605 (2) & 0.005 (2) &  & 0.373 (2) & 0.001 (2) &  \\
MicroBooNE all & 0.114 (6) & 0.556 (6) & 0.016 (6) & 0.473 (5) & 0.398 (6) & 0.004 (6) & 0.622 (5) \\
T2K+MicroBooNE & 0.049 (4) & 0.740 (3) & 0.000 (3) & 0.000 (2) & 0.018 (4) & 0.003 (4) & 0.772 (2) \\
T2K+MicroBooNE all & 0.099 (8) & 0.605 (7) & 0.000 (7) & 0.000 (6) & 0.037 (8) & 0.006 (8) & 0.678 (6) \\
all & 0.000 (10) & 0.000 (9) & 0.000 (9) & 0.000 (7) & 0.001 (9) & 0.000 (9) & 0.011 (7) \\
\bottomrule
    \end{tabular*}
\end{table*}

All $\fmax$ test statistics described above are implemented in \cite{Koch2024}.
Since the differences between the $\pmin$ and $\fmaxopt$ statistics are rather small,
the former seems to be a very good candidate to do ``quick and dirty'' combinations of separate model checks.
For small p-values, this is as simple as multiplying the lowest p-value by the number of data blocks.

\section{Parameter estimation in the presence of unknown correlations}
\label{sec:derate}

The previously described test statistics (here and in \cite{Koch2021}) are suitable for simple hypothesis tests, i.e. checking if a particular model without any free parameters is compatible with the data.
But they have some disadvantages when it comes to parameter estimation, i.e. fitting variable parameters of a model to the data:
\begin{itemize}
    \item They are not smoothly differentiable everywhere.
    \item They can have multiple local minima.
    \item There is no equivalent to Wilks' theorem, and it is not clear how the difference to the best fit value should be distributed.
\end{itemize}
Especially the last point makes it difficult to construct confidence intervals for the model parameter values.
Of course, it is always possible to do the point-wise inclusion/exclusion for all points in the available model space to construct a confidence region that way (forgoing Wilks' theorem),
but usually when doing parameter estimation it is desirable to guarantee that there will be a non-empty confidence region.
This is especially the case when determining parameters of effective models,
which are unlikely to actually perfectly reflect the true physical processes and data expectation values,
and even their ``best fit point'' is excluded at the desired confidence level.
So, to facilitate parameter estimations a different approach should be chosen.

If we ignore the possible correlations, the obvious choice for a parameter estimation would be using the naive M-distance.
It is smooth, has a unique minimum in any linear subspace, and the expected distribution for the difference between the fitted minimum value at $\bm\mu(\bm{\hat\theta})$ and the M-distance at the true point $\bm\mu(\bm{\theta}_0)$ is a chi-squared distribution with as many degrees of freedom as fit parameters $k$:
\begin{widetext}
\begin{align}
    \qty((\bm\mu(\bm{\theta}_0) - \bm{x})^T S^{-1} (\bm\mu(\bm{\theta}_0) - \bm{x}))
    - \qty((\bm\mu(\bm{\hat\theta}) - \bm{x})^T S^{-1} (\bm\mu(\bm{\hat\theta}) - \bm{x}))
    \sim \chi^2_k.
\end{align}
\end{widetext}

To investigate how this setup behaves in the presence of unknown correlations, we will use that a parameter fit, to the first approximation, is equivalent to a projection onto a linear subspace.
The naive test statistic in our toy example is then:
\begin{align}
    \naive(\bm{x} | \bm\mu, S_0) =&  (\bm\mu - \bm{x})^T S_0^{-1} (\bm\mu - \bm{x}) ,\\
    S_0 =& I_{10} .
\end{align}
This is representative for the general case, where the known covariance blocks $S_{11}$ and $S_{22}$ are \emph{not} identity matrices,
as all measurements can be brought into this form by a linear variable transformation.

Now let us assume a parameterization of a model that describes the expectation values for the data in the form
\begin{align}
    \bm\mu(\bm\theta) = \bm{x}_0 + A \bm\theta.
\end{align}
Here $\bm\theta$ is a vector of $k$ model parameters and $A$ is a matrix of $k$ column vectors that span the linear subspace that is available to the model.
Again, for notation simplicity, we will assume that $\bm{x}_0 = \bm\mu(\bm{0})$ is the true expectation value that the data is distributed around:
\begin{align}
    \bm{x} \sim \normal(\bm\mu = \bm{x}_0, \Sigma=V)
\end{align}
A fit to the data will find the estimate $\bm{\hat\theta}$ which minimizes the test statistic:
\begin{align}
    \bm{\hat\theta} &= \argmin_{\bm\theta} \naive(\bm\mu(\bm\theta)|\bm{x}, S_0), \\
    \bm{\hat{x}} &= \bm\mu(\bm{\hat\theta}),
\end{align}
which, for this linear model, is equivalent to a projection (see e.g. \cite[Sec.\,19.1]{Harville1997}):
\begin{align}
    \bm{\hat\theta} &= Q(\bm{x} - \bm{x}_0), \\
    Q &= (A^TS_0^{-1}A)^{-1}A^TS_0^{-1}, \\
    \bm{\hat{x}} &= \bm\mu(\bm{\hat\theta}) = P(\bm{x} - \bm{x}_0) + \bm{x}_0, \\
    P &= AQ.
\end{align}
This linear approximation can also be made for non-linear models around the best fit point:
\begin{align}
    \bm\mu(\bm\theta) &= \bm{f}(\bm\theta) \approx \bm{\hat{x}} + A(\bm\theta - \bm{\hat\theta})), \\
    A^T &= \nabla_{\bm\theta} \bm{f}^T(\bm{\hat\theta}),
\end{align}
where $A$ is the Jacbobian matrix of $\bm{f}$ wrt. $\bm\theta$ at $\bm{\hat\theta}$.

With the optimization problem cast as a projection, it is now simple to calculate the
expected distribution of $\bm{\hat\theta}$:
\begin{align}
    \bm{\hat\theta} &\sim \normal(\bm\mu = \bm{0}, \Sigma=V_\theta) ,\\
    V_\theta &= QVQ^T.
\end{align}
Assuming that $V = S_0$, i.e. that there actually are no unaccounted correlations, we can define the naive test statistic for the parameter estimation:
\begin{widetext}
\begin{align}
    S_{\theta0} &= QS_0Q^T = (A^TS_0^{-1}A)^{-1}A^TS_0^{-1} S_0 ((A^TS_0^{-1}A)^{-1}A^TS_0^{-1})^T \nonumber\\
             &= (A^TS_0^{-1}A)^{-1}A^T S_0^{-1} A ((A^TS_0^{-1}A)^{-1})^T \nonumber\\
             &= ((A^TS_0^{-1}A)^{-1})^T = (A^TS_0^{-1}A)^{-1},\\
    \naive(\bm{\hat\theta} | \bm\theta, S_{\theta0}) &= (\bm\theta - \bm{\hat\theta})^T S_{\theta0}^{-1} (\bm\theta - \bm{\hat\theta}). \label{eq:parameters}
\end{align}
\end{widetext}
This will be $\chi^2_k$ distributed for the true parameter value $\bm\theta = \bm\theta_0 = \bm{0}$ if $V = V_0 = S_0$ and thus $V_\theta = V_{\theta 0} = S_{\theta 0}$.

When there are unaccounted correlations present, the distribution will look different.
The test statistic is a quadratic form of a multivariate normal distribution,
so in general it will be distributed like a ``generalised chi-square distribution'' $\tilde\chi^2_{\bm{w}, \bm{k}, \bm{\lambda}, s, m}$\cite{Das2020}.
Since the distribution of $\bm{\hat\theta}$ is centred around the true value $\bm{0}$,
we can reduce the complexity of the distribution a bit.
Following the steps in \cite{Das2020}:
\begin{widetext}
\begin{align}
    \bm{q}(\bm{\hat\theta})
        &= \naive(\bm{\hat\theta}|\bm{0}, S_{\theta0}) = \bm{\hat\theta}^T S_{\theta0}^{-1} \bm{\hat\theta} ,\\
    \bm{z} &= V_\theta^{-1/2} \bm{\hat\theta} && \sim \normal(\bm\mu = \bm{0}, \Sigma=I_k) ,\\
    \bm{q}(\bm{z}) &= \bm{z}^T \tilde{S}_{\theta0}^{-1} \bm{z} ,\\
    \tilde{S}_{\theta0}^{-1} &= V_\theta^{1/2} {S}_{\theta0}^{-1} V_\theta^{1/2} = RDR^T,\\
    \bm{y} &= R^T \bm{z} && \sim \normal(\bm\mu = \bm{0}, \Sigma=I_k) ,\\
    \bm{q}(\bm{y}) &= \bm{y}^T D \bm{y} = \sum_i^k d_{i} y_i^2 && \sim \tilde\chi^2_{\bm{d}, \bm{1}, \bm{0}, 0, 0} \label{eq:fit-dist},
\end{align}
\end{widetext}
where $V_\theta^{-1/2}$ denotes the inverse of the symmetric square root of $V_\theta$.
$RDR^T$ is the eigen value decomposition of $\tilde{S}_{\theta0}^{-1}$,
with an orthogonal $R$ and $\bm{d}$ the vector of eigenvalues making up the diagonal matrix $D$.
So the naive test statistic is distributed like the weighted sum of $k$ indipendent, $\chi^2_1$ distributed random variables,
with the weights corresponding to the eigenvalues of $\tilde{S}_{\theta0}^{-1}$.
If $V = S_0$, then $\tilde{S}_{\theta0}^{-1} = I_k$ and the distribution becomes $\chi^2_k$, as expected.

\begin{figure*}
    \centering
    \includegraphics[width=0.49\textwidth]{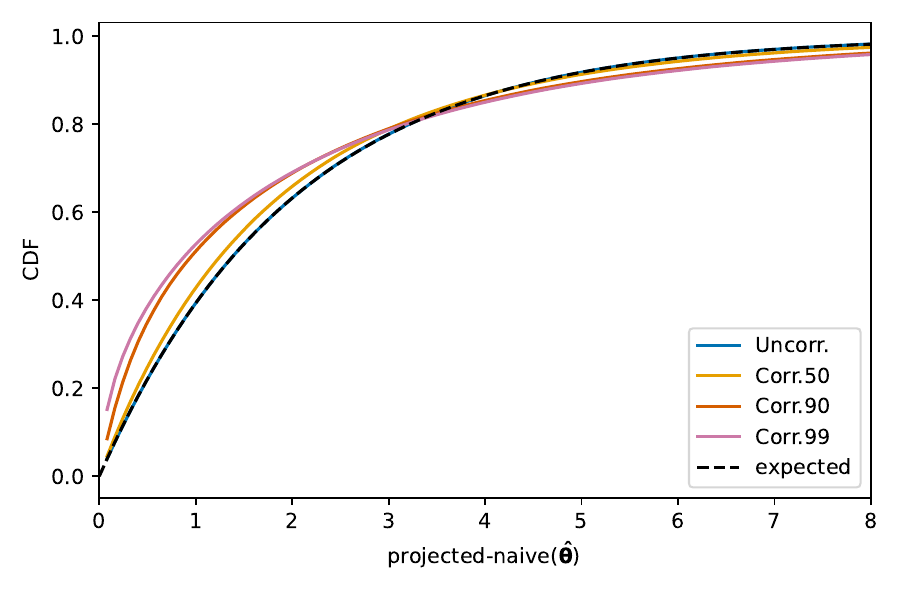}
    \includegraphics[width=0.49\textwidth]{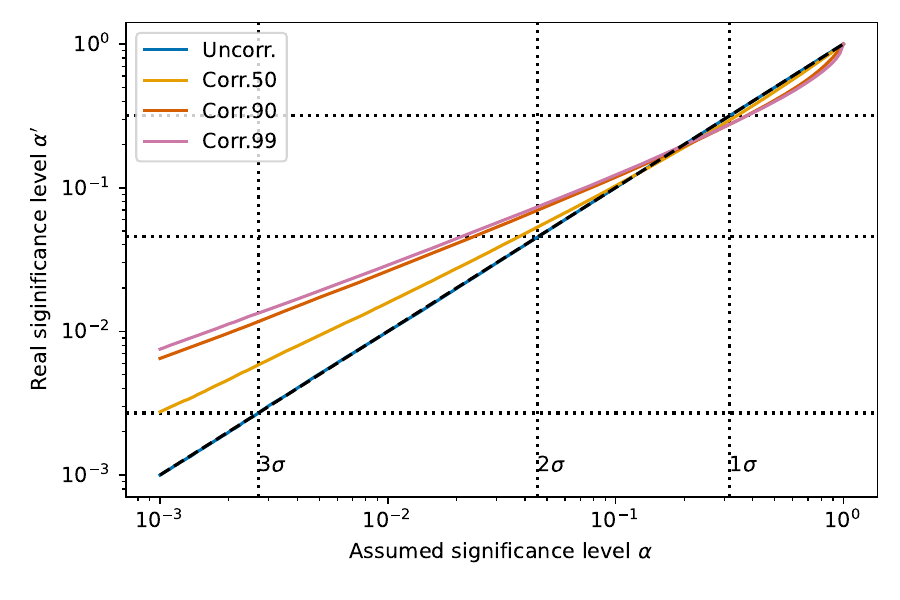}
    \caption{\label{fig:projected-naive}%
        CDFs (left) for the ``naive'' squared M-distance in the projected parameter space for different levels of correlations in the data.
        This is equivalent to a parameter estimation by running a fit.
        When using the uncorrelated CDF to calculate the assumed significance level (or p-value) of a value of the statistic, the actual level will differ from the assumption depending on the correlations (right).
        Where the real significance level is larger than the assumed significance level (the real significance is weaker than the assumed one), the test statistic shows undercoverage.
        This corresponds to the region in the CDF where the real CDF for a given value is lower than the expected one.
    }
\end{figure*}

To visualise this, we will use the same toy data as before.
We implement a 2D fit by setting
\begin{gather}
    A = \mqty*(9 & 8 & 7 & 6 & 5 & 4 & 3 & 2 & 1 & 0 \\
               1 & -1 & 1 & -1 & 1 & -1 & 1 & -1 & 1 & -1)^T.
\end{gather}
The result can be seen in \autoref{fig:projected-naive}.
Like in the simple hypothesis test case shown in \autoref{fig:block-naive},
the true significance of a test statistic value deviates from the assumed significance when there are unaccounted correlations present.
Note, that in this case, the effect depends on the subspace spanned by $A$ and how it relates to the actual correlations.
E.g. if $A$ consists only of vectors with non-zero elements in the first block of variables, but not the second block, none of the correlations introduced between the blocks will matter,
and the naive M-distance will always be exact.
This is a trivial example, as it would mean the model parameters do only affect one of the two data sets, so the second data set might as well be ignored.
But this can also happen in less obvious ways, depending on how the two blocks are actually correlated to one another, which is unknown in the problem discussed here.

In general $V_{\theta}$ no longer has the sharp distinction between known and unknown covariance elements that was present in $V$, and thus it is generally not possible to simply apply the previously developed robust test statistics to the model parameter space.
Instead, in order to ensure that the coverage is conservative at least up to a set significance, e.g. $3\sigma$, we can inflate the uncertainties by a constant factor.
This will scale the values of the test statistic accordingly and thus also the CDF.

\begin{figure*}
    \centering
    \includegraphics[width=0.49\textwidth]{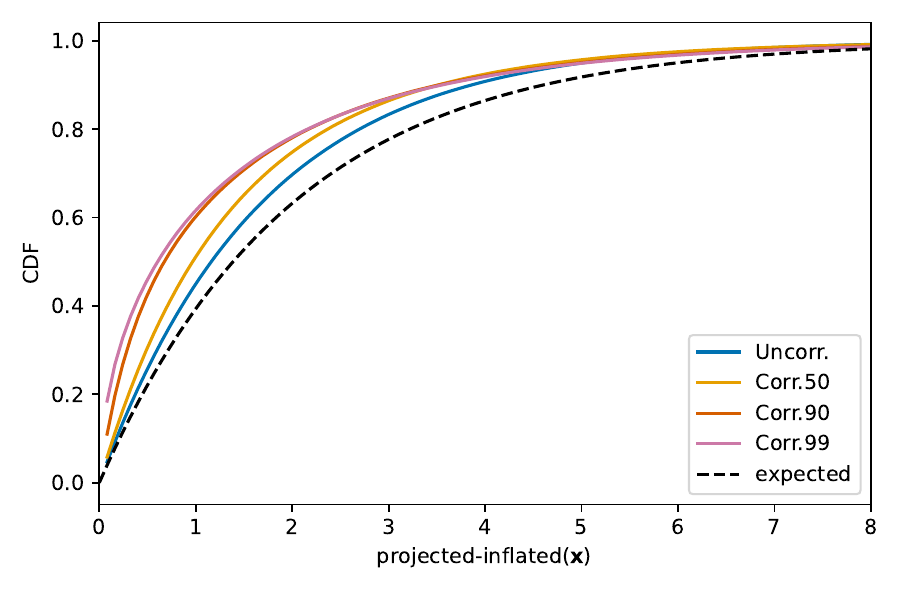}
    \includegraphics[width=0.49\textwidth]{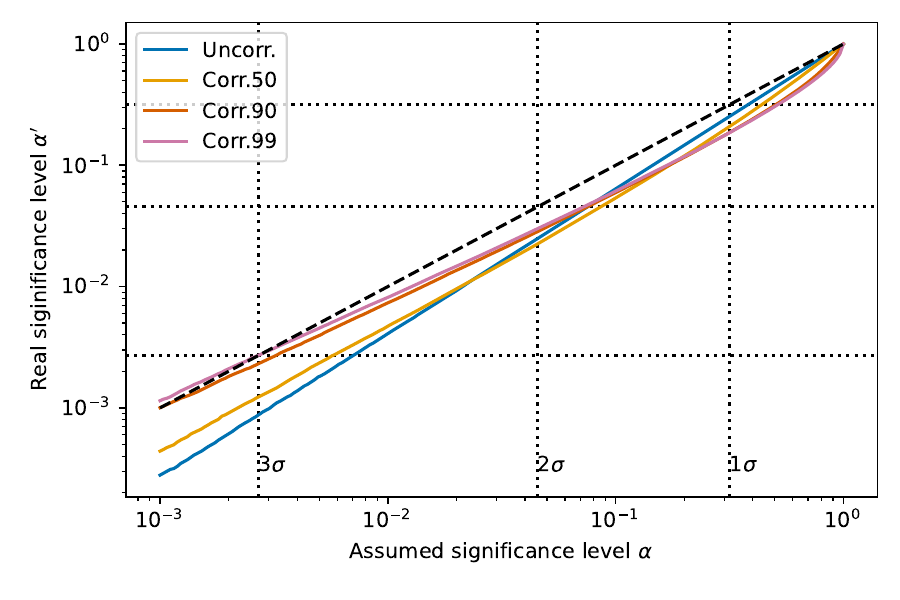}
    \caption{\label{fig:projected-inflated}%
        CDFs (left) for the inflated squared M-distance in the projected parameter space for different levels of correlations in the data.
        This is equivalent to a parameter estimation by running a fit.
        Compared to \autoref{fig:projected-naive}, the assumed covariance is inflated by a factor $\alpha = 1.20$.
        Now the test statistic remains conservative even for the strongest correlation investigated in the toy data, up to a significance of $3\sigma$.
    }
\end{figure*}

\autoref{fig:projected-inflated} shows how the coverage behaves if the assumed covariance is inflated by a factor $\alpha = 1.20$:
\begin{align}
    \inflated(\bm{\hat\theta}|\bm{0}, S_{\theta0}, \alpha) = \frac{\bm{\hat\theta}^T S_{\theta0}^{-1} \bm{\hat\theta}}{\alpha}.
\end{align}
This value of $\alpha$ was determined by creating a separate toy data set with a covariance $V_1$ of the form in \autoref{eq:toy-cov}, but with the strongest possible correlation $\rho = 1$.
From this toy data one calculates the distribution of $\naive(\bm{\hat\theta})$ and then determines the Inverse Distribution Function (IDF, or quantile function) $F^{-1}$.
The necessary scale factor is then just the ratio of this IDF and and the expected IDF of the expected $\chi^2$ distribution when $V=V_0=S_0$, at the desired confidence level $\gamma$:
\begin{align}
    \gamma &= 0.997 ,\\
    \alpha &= \frac{F^{-1}_{V_1}(\gamma)}{ F^{-1}_{V_0}(\gamma)} ,\\
    F_{V_0} &= F_{\chi^2_k}.
\end{align}

This factor however is too small to ensure conservativeness up to $3\sigma$ for \emph{all} possible correlations between the two blocks of data.
To get that factor, one in principle needs to maximise the IDF over all possible covariances:
\begin{align}
    \alpha_{\max} = \max_V \frac{F^{-1}_{V}(\gamma)}{ F^{-1}_{V_0}(\gamma)} ,
\end{align}
where the maximisation over V can only affect the unknown covariance elements and must ensure that $V$ remains positive semi definite.
This is possible, but challenging on a technical level.

\section{Algorithmic determination of the scaling factor}
\label{sec:algo}

We will now introduce an approximate way to determine the scaling factor $\alpha$,
which does not guarantee to find $\alpha_{\max}$, but which seems to perform well in practice.
First, the known covariance blocks $S_{ii}$ have to be brought into standard normal form.
This can be done by a block-diagonal whitening transform:
\begin{align}
    W &= \mqty*( W_{11} &        & 0 \\
                        & W_{22} &   \\
                      0 &        & \ddots
                ) ,\\
    \bm\xi &= W \bm{x} \hspace{4em} \sim \normal(\bm\mu = 0, \Sigma=V_\xi) ,\\
    V_\xi &= WVW^T = \mqty(I_{N_1} & V_{\xi12}    & V_{\xi13} &  \\
                     V^{T}_{\xi12} & I_{N_2}     & V_{\xi23} & \cdots \\
                     V^{T}_{\xi13} & V^{T}_{\xi23} & I_{N_3}  &  \\
                                 & \vdots      &          & \ddots{} ) \text{,}
\end{align}
with suitable block-wise whitening matrices, so $W_{ii}^T W_{ii} = S_{ii}^{-1}$.
In this coordinate system, the naively expected covariance matrix $S_{0 \xi} = W S_0 W^T$ is equal to the identity matrix $I_N$,
and the Jacobian matrix $A_\xi = W A$.
We will later discuss which whitening transforms are best suited in this application.

From \autoref{eq:fit-dist} we know the expected distribution in terms of a weighted sum of $\chi^2$ distributed variables.
These weights are the eigenvalues of the matrix $\tilde{S}_{\theta0}^{-1}$,
so we can express the expected value and variance of that distribution in terms of matrix traces:
\begin{widetext}
\begin{align}
    E[\naive(\bm{\hat\theta}|\bm{0}, S_{\theta 0})] &= \sum_i d_i
      = \Tr(\tilde{S}_{\theta0}^{-1}) \nonumber\\
      &= \Tr(V_\theta^{1/2} S_{\theta 0}^{-1}  V_\theta^{1/2}) = \Tr(V_\theta S_{\theta 0}^{-1}) \nonumber\\
      &= \Tr(Q_\xi V_\xi Q_\xi^T A_\xi^TS_{0 \xi}^{-1} A_\xi) \nonumber\\
      &= \Tr((A_\xi^T A_\xi)^{-1}A_\xi^T V_\xi A_\xi (A_\xi^T A_\xi)^{-1 T} A_\xi^T A_\xi) \nonumber\\
      &= \Tr( A_\xi (A_\xi^T A_\xi)^{-1}A_\xi^T V_\xi ) =  \Tr( P_\xi V_\xi ) ,\\
    \Var[\naive(\bm{\hat\theta}|\bm{0}, S_{\theta 0})] &= \sum_i 2 d_i^2
      = 2 \Tr((\tilde{S}_{\theta0}^{-1})^2)
      = 2 \Tr(P_\xi V_\xi P_\xi V_\xi),
\end{align}
\end{widetext}
with the projection matrices $Q_\xi = (A_\xi^T S^{-1}_{0\xi} A_\xi)^{-1}A_\xi^T S^{-1}_{0\xi} = (A_\xi^T A_\xi)^{-1}A_\xi^T$ since $S_{0 \xi} = I_N$, and $P_\xi = A_\xi Q_\xi$.
Now instead of trying to maximize the IDF directly, we can try to find off-diagonal elements for $V_\xi$ that maximize the expected value and variance of the distribution,
to shift the value of the IDF for a given confidence level to as high values as possible.

Since $V_\xi$ and $P_\xi$ are symmetric, we have (see e.g. \cite[Sec.\,15.6]{Harville1997})
\begin{align}
    \dv{V_\xi} E[\naive] = \dv{V_\xi} \Tr(P_\xi V_\xi) &= P_\xi ,\\
    \dv{V_\xi} \Var[\naive] = 2 \dv{V_\xi} \Tr(P_\xi V_\xi P_\xi V_\xi) &= 4 P_\xi V_\xi P_\xi .
\end{align}
If $P_\xi = I_N$, which corresponds to a fit with $N$ free parameters or simply using the data for simple hypothesis tests,
only the diagonal elements of $V_\xi$ contribute to the expected value.
These are perfectly known and thus the expected value is equal to $N$.
Since the sum of eigenvalues is constant, the variance is maximized by ``concentrating'' this sum in as few eigenvalues as possible,
i.e. set as many eigenvalues as possible to $0$.
This is achieved by setting as many elements of $V_\xi$ to $\pm 1$ as possible.
Before setting any elements to $\pm 1$, $V_\xi = I_N$ and $4 P_\xi V_\xi P_\xi = 4 P_\xi$.
This means we gain the most variance by choosing the off-diagonal elements of $V_\xi$ that correspond to the largest absolute values of $P_\xi$.

In general, for any given sum of eigenvalues $\Tr(P_\xi V_\xi)$, the sum of squared eigenvalues $\Tr(P_\xi V_\xi P_\xi V_\xi)$ is maximized by ``concentrating'' the former trace in as few eigenvalues of $P_\xi V_\xi$ as possible.
The number of non-zero eigenvalues is limited by the rank of a matrix.
The rank of a matrix product is limited by the smallest rank of the matrices.
The rank of $V_\xi$ is minimized by setting as many elements to $\pm 1$ as possible.
In order to maximize $\Tr(P_\xi V_\xi)$ while minimizing the rank of $V_\xi$,
the following algorithm has proven useful:
\begin{enumerate}
    \item \label{itm:start} Among the free elements of $V_\xi$, select the one with the largest contribution to $\Tr(P_\xi V_\xi)$:
        \begin{align}
            (i,j) = \argmax_{(i,j) \in \text{free}} | P_{\xi ij} |
        \end{align}
    \item Set the element to $\pm 1$ according to the sign of the derivative:
        \begin{align}
            V_{\xi ij} = \sign( P_{\xi ij} ),
        \end{align}
        where $\sign(0) = 1$ is defined as a special case.
    \item Depending on the set elements in $V_\xi$ this newly set value also forces some other elements to be a certain value. Set them accordingly.
    \item If free elements remain, repeat from \autoref{itm:start}.
\end{enumerate}
A Python implementation for this algorithm is available as part of the NuStatTools Python package\cite{Koch2024}.

\section{Choice of whitening transform}

There are many possible choices of whitening transforms for the $W_{ii}$ matrices.
In fact, there are infinitely many whitening transforms, as one can always multiply a whitening matrix $W'$ with an orthogonal matrix $R$ to get a new whitening transform,
\begin{align}
    W &= R W', \\
    W^TW &= W'^T R^T R W' = W'^T W' = S^{-1}.
\end{align}
For this algorithm, we would like to maximize the impact that each off-diagonal $\pm1$ in the correlation matrix has on the distribution of the test statistic.
This is the case if the base vectors in the whitened space are embedded in the image space of the projection matrix $P_\xi$.
This way, the correlation between those base vectors has the maximum effect in the fit parameter space.
If this is not immediately obvious, it is useful to think about the influence of correlations between base vectors that are orthogonal to the image space of the Projection matrix $P_\xi$.
These vectors have no influence on the result of the fit, so their correlations will also have absolutely no influence on the distribution of the best fit points, and thus on the test statistic.

The projection matrix in the whitened coordinates is
\begin{align}
    P_\xi &= R W' P W'^{-1} R^T.
\end{align}
To align a base vector of this space with the image space of $P_\xi$,
the columns of $R^T$ should be eigenvectors of $W' P W'^{-1}$.
Since we only have the freedom to choose diagonal block matrices $R_{ii}$, this is not always possible.
Instead we construct columns in $R^T$ that are ``close enough'':
\begin{align}
    W' P W'^{-1} &= \mqty(\xmat*{P'}{3}{3} & \\ & & & \ddots) ,\\
    P_{ii}' &= U_{ii} D_{ii} U_{ii}^T ,\\
    R &= \mqty(\dmat[0]{R_{11}, R_{22}, R_{33}} & \\ & & & \ddots) ,\\
    R_{ii} &= U_{ii}^T ,
\end{align}
where $U_{ii}D_{ii}U_{ii}^T$ is the singular value decomposition of $P'_{ii}$.
The choice of the intermediate whitening transform is not important and we can choose e.g. $W'_{ii} = S_{ii}^{-1/2}$.

\begin{figure*}
    \centering
    \includegraphics[width=0.49\textwidth]{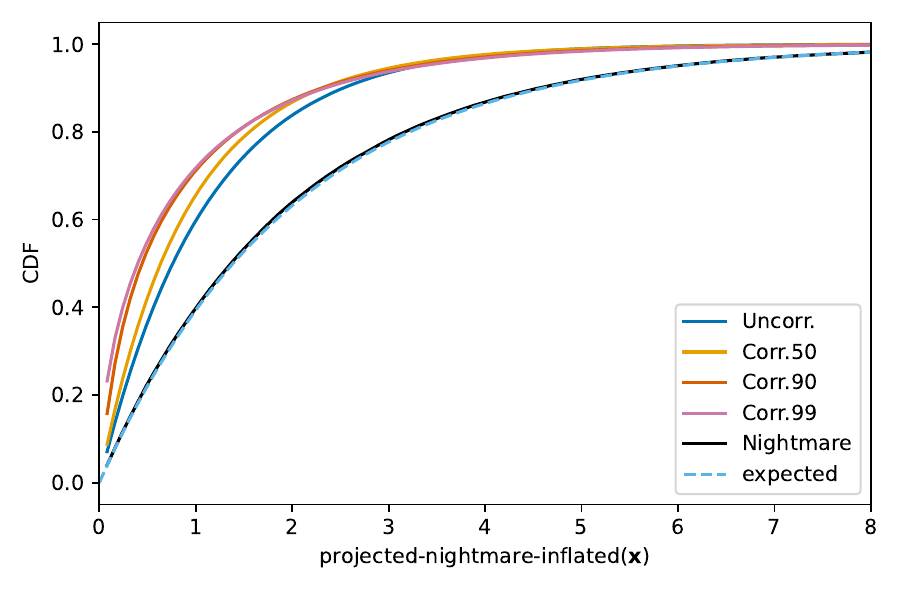}
    \includegraphics[width=0.49\textwidth]{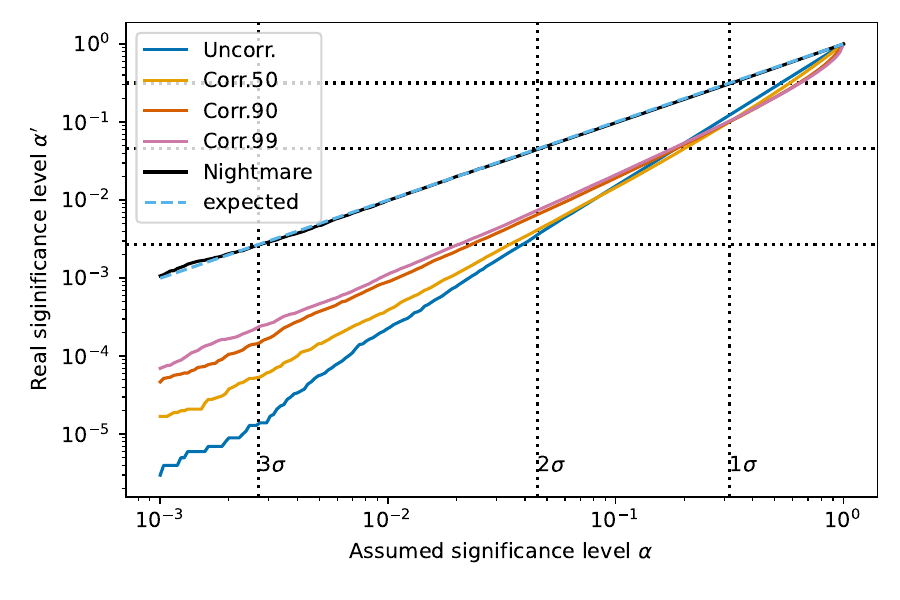}
    \caption{\label{fig:projected-nightmare-inflated}%
        CDFs (left) for the inflated squared M-distance in the projected parameter space for different levels of correlations in the data.
        This is equivalent to a parameter estimation by running a fit.
        Compared to \autoref{fig:projected-naive}, the assumed covariance is inflated by a factor $\alpha = 1.82$.
        This factor ensures conservative coverage even for the possible ``nightmare'' covariance, up to a significance of $3\sigma$.
        The correlation in the toy data sets has a weaker impact and thus there is still over-coverage even at the strongest level of correlation.
        The differences in distribution between the inflated test statistic and the expected $\chi^2_2$ are very small for the ``nightmare'' data in this example.
        This is not always the case.
    }
\end{figure*}

\begin{figure}
    \centering
    \includegraphics[width=0.49\textwidth]{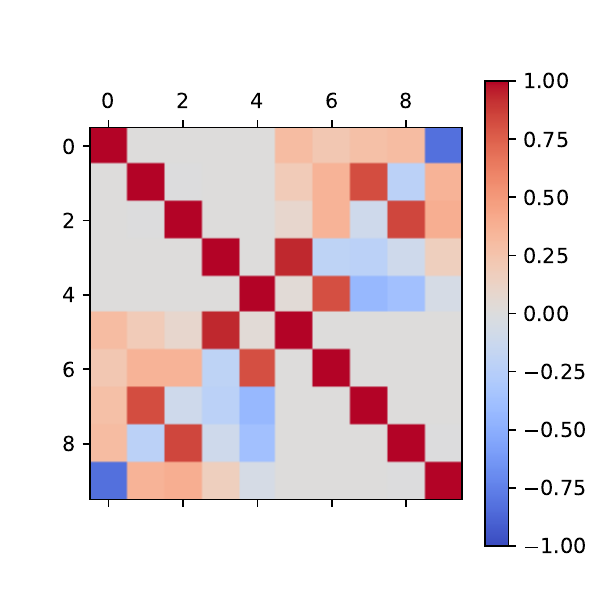}
    \includegraphics[width=0.49\textwidth]{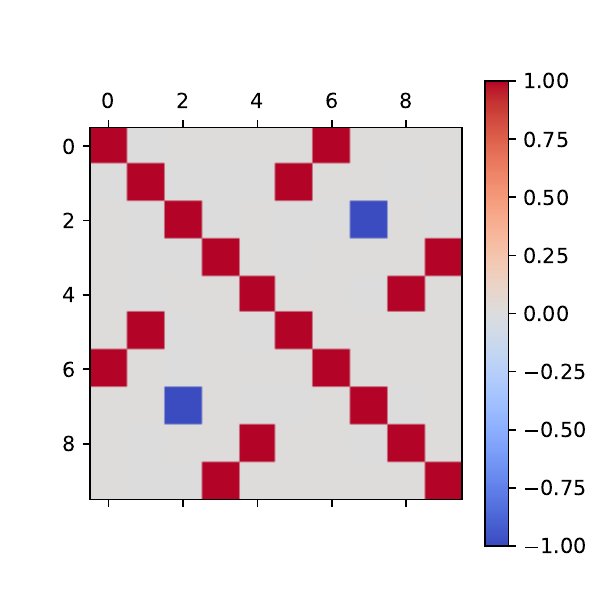}
    \caption{\label{fig:nightmare-cov}%
        The ``nightmare'' covariance in the original coordinate system $V_\ddagger$ (top) and in the aligned whitened coordinate system $V_{\xi\ddagger}$ (bottom) for the toy data as determined by the algorithm described in the main text.
    }
\end{figure}

By applying this algorithm we can construct a ``nightmare'' covariance $V_{\xi\ddagger}$,
which we then use to calculate the scaling factor $\alpha_\ddagger$:
\begin{align}
    \alpha_\ddagger &= \frac{F^{-1}_{V_{\xi\ddagger}}(\gamma)}{ F^{-1}_{V_{\xi 0}}(\gamma)}. \label{eq:derate}
\end{align}
Applied to the previous toy example, this yields $\alpha_\ddagger = 1.82$,
a worse inflation than what was needed to cover the correlations actually present in the dataset, as shown in \autoref{fig:projected-nightmare-inflated}.
This is due to how the model space is oriented wrt. the actual correlations in the data.
Note that for the toy example $V_{\xi\ddagger}$ as shown in \autoref{fig:nightmare-cov} has some negative elements while the actual covariance is purely positive.
Since the true correlations are unknown in real problems of the type discussed in this paper, we have no choice but to assume the worst.

\section{Approximate derating factor for simple hypothesis tests}

The derating factor depends non-trivially on the details of the size of the data blocks, and how the Jacobian of the fitted model relates to the uncertainties.
This makes it difficult to make general statements about the size of the derating factor.
But we can make some statements about the case where $A = I$, i.e. when the covariance is used for simple hypothesis tests.
As mentioned before, in this case the expected value of the $\naive$ test statistic is constant and equal to the total number of bins.
Let $N_i$ be the number of bins in the $i$th of the $N$ blocks, and we now order things such that $N_i$ are decreasing with increasing $i$, so $N_1$ is the largest and $N_N$ is the smallest block.
By applying the algorithm above, all $N_N$ bins in the smallest block get 100\% correlated to bins in all other blocks.
This determines the largest eigenvalue, $N$, with a multiplicity of $N_N$.
At the same time, $N_N$ bins in all larger blocks are ``used up'' and only contribute as eigenvalues of $0$.
Only the remaining bins of the next larger block $N_{N-1}$ then get correlated with the blocks $N-2$ and larger, contributing the eigenvalue $N-1$ with multiplicity $N_{N-1} - N_{N}$, and so on.
In general the eigenvalue $i \in \{1, \dots, N\}$ has a multiplicity of $N_{i} - N_{i+1}$ with $N_{N+1} = 0$.
So the variance is
\begin{align}
    \Var[\naive] &= 2 \sum_{i=1}^N i^2 (N_i - N_{i+1}) \nonumber \\
        &= 2 \qty( \qty(\sum_{i=1}^N i^2 N_i) - \qty(\sum_{j=2}^N (j-1)^2 N_j )) \nonumber \\
        &= 2 \qty( \qty(\sum_{i=1}^N 2 i N_i) - \qty(\sum_{j=1}^N N_j )) \nonumber \\
        &= 2 \qty( \frac{\sum_{i=1}^N 2 i N_i}{\sum_{j=1}^N N_j} - 1) \sum_{j=1}^N N_j  \nonumber \\
        &= 2 \qty(2 \ibar - 1) k,
\end{align}
with the total number of bins $k$ and the average block number $\ibar$ over all bins.
For equally sized blocks, the latter will be equal to $(N+1)/2$,
while for unequal lock sizes it will be smaller.

\begin{figure}
    \centering
    \includegraphics[width=0.49\textwidth]{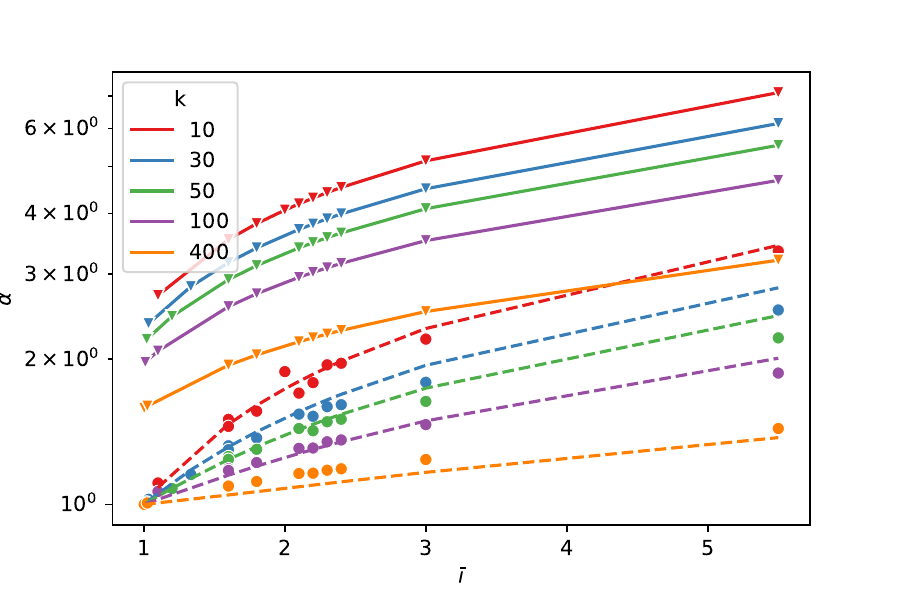}
    \includegraphics[width=0.49\textwidth]{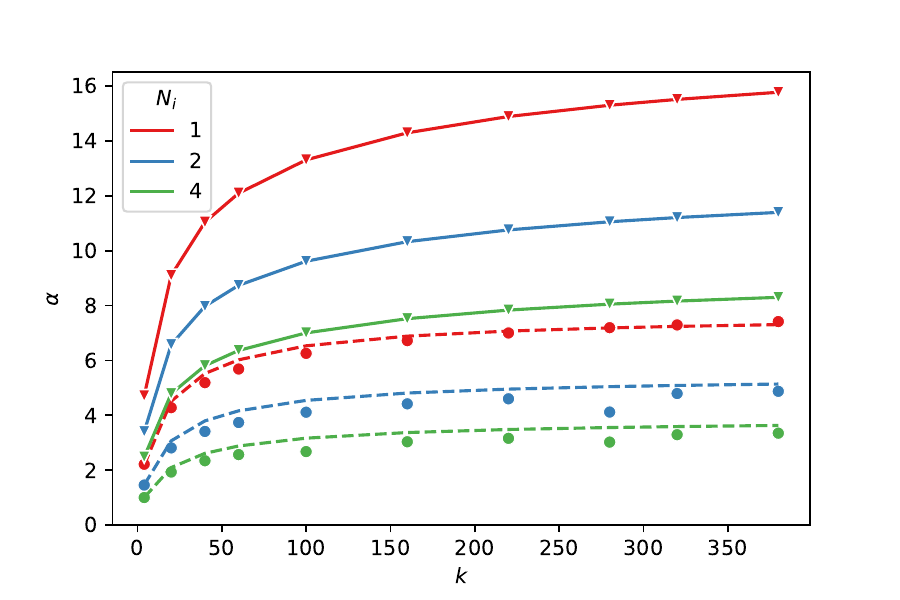}
    \caption{Derating factors for various total number of bins $k$ and average block number $\bar{i}$ (top) or constant block size $N_i$ (bottom) combinations.
    The data points show the algorithmically determined ``nightmare'' derating factors.
    Solid lines show the upper limit, and the dashed lines the approximation as described in the main text.
    A constant block size of $N_i = 1$ means that we are dealing with $k$ data points with completely unknown correlations between any of them.
    In all cases here, the target conservative confidence level is $\gamma=0.997$.}
    \label{fig:approx}
\end{figure}

Using the Vysochanskij–Petunin inequality\cite{Mercadier2021}, we can construct bounds for the CDF and IDF:
\begin{align}
    P(X > E[X] + r) &\le \frac{4}{9} \frac{\Var[X]}{r^2 + \Var[X]} \nonumber\\
    &\qfor r^2 > \frac{5}{3} \Var[X],
\end{align}
\begin{align}
    F_X(x = E[X]+r) &\ge 1 - \frac{4}{9} \frac{\Var[X]}{(x - E[X])^2 + \Var[X]},
\end{align}
\begin{align}
    F^{-1}_X(\gamma) &\le \sqrt{\Var[X]\qty(\frac{4}{9(1-\gamma)} - 1)} + E[X].
\end{align}
Thus the derating factor is limited by
\begin{align}
    \alpha_\ddagger &\le \frac{\sqrt{\Var[X]\qty(\frac{4}{9(1-\gamma)} - 1)} + E[X]}{F^{-1}_{V_{\xi0}}(\gamma)} \\
        &= \frac{\sqrt{2 \qty(2 \ibar - 1) k\qty(\frac{4}{9(1-\gamma)} - 1)} + k}{F^{-1}_{V_{\xi0}}(\gamma)}.
\end{align}
\autoref{fig:approx} shows these limits compared to actually computed derating factors for a target conservative significance up to $3\sigma$, i.e. $\gamma = 0.997$.
The limits are not very strong.
More useful is an approximation of the derating factor:
\begin{align}
    \alpha_\approx &= \sqrt{1 + 120 \frac{\ibar - \sqrt{\ibar}}{k + 25}},
\end{align}
which was determined using symbolic regression\cite{Cranmer2023} on the calculated derating factors.
For a wide range of possible combinations of data block sizes, the derating factor is actually smaller than the proposed doubling of the variance as discussed in \cite{Meng2022}.

\section{Presence of some known off-diagonal covariance blocks}

So far we have assumed that the diagonal blocks of the covariance are known and the off-diagonal blocks are all unknown.
It might be reasonable to assume that some of the off diagonal blocks are known to be 0, though.
A possible example would be a fit to multiple results from multiple experiments.
The results of a single experiment probably have some correlations, like shared systematic uncertainties,
while it could be justified to assume that two very dissimilar experiments are perfectly uncorrelated.
In this case, the off-diagonal blocks of $V$ between results of different experiments can be assumed to be $V_{ij} = S_{0ij} = 0$.

The algorithm described above can still be applied in these cases.
The off-diagonal blocks of the whitened covariance are $V_{\xi ij} = W_{ii}V_{ij}W_{jj} = 0$,
so they do not depend on the unknown covariances.
They will just be set to 0 and not touched by the algorithm.

\section{Mix of known and unknown contributions to the covariance}

It is not unusual to have some information about the possible correlations.
For example, it might be possible to split the covariance of two experiments into a statistical and systematic contribution.
The statistical parts should then be perfectly uncorrelated, while the correlations of the systematic uncertainties are unknown.
Or the systematic uncertainties can be split up into parts that are definitely uncorrelated, like uncertainties about the detector performance of the two experiments, and those that are potentially correlated, like physics model uncertainties.

If it is possible to split the total covariance into separate parts like this, it is possible to generate the ``nightmare scenario'' for each of them separately:
\begin{align}
    V &= \sum_i V_i, \\
    V_{i \xi_i} &= W_i V_i W_i^T, \\
    V_\ddagger &= \sum_i W_i^{-1} V_{i \xi_i \ddagger} W_i^{-1T}.
\end{align}
For this, the blocks of known correlations do not have to be the same size for the separate $V_i$ and also the blockwise whitening transforms $W_i$ will be different.
The $\xi_i$ indices denote matrices in the coordinate system where $V_i$ is whitened.
Each single $V_{i \xi_i \ddagger}$ is constructed similarly as described in \autoref{sec:algo}.

The aim is again to calculate the ``nightmare covariances'' which ensure the highest possible expectation value and variance for the naive test statistic:
\begin{widetext}
\begin{align}
    E[\naive(\bm{\hat\theta}|\bm{0}, S_{\theta 0})] &= \sum_i d_i
        = \Tr(\tilde{S}_{\theta0}^{-1})
        = \Tr( V_\theta S_{\theta 0}^{-1}) \nonumber\\
        &= \sum_i \Tr( V_{\theta i} S_{\theta 0}^{-1}) \nonumber\\
        &= \sum_i \Tr(Q_{\xi_i} V_{i \xi_i} Q_{\xi_i}^T A_{\xi_i}^TS_{0 \xi_i}^{-1} A_{\xi_i}) \nonumber\\
        &= \sum_i \Tr((A_{\xi_i}^T S^{-1}_{0 \xi_i} A_{\xi_i})^{-1}A_{\xi_i}^T S^{-1}_{0 \xi_i}
        V_{i \xi_i}
        S^{-1T}_{0 \xi_i}A_{\xi_i}(A_{\xi_i}^T S^{-1}_{0 \xi_i} A_{\xi_i})^{-1T}
        A_{\xi_i}^TS_{0 \xi_i}^{-1} A_{\xi_i}) \nonumber\\
        &= \sum_i \Tr(V_{i \xi_i}
        S^{-1T}_{0 \xi_i}A_{\xi_i} Q_{\xi_i})
        = \sum_i Tr(V_{i \xi_i} T_i),\\
    \Var[\naive(\bm{\hat\theta}|\bm{0}, S_{\theta 0})] &= \sum_i 2 d_i^2
        = 2 \Tr((\tilde{S}_{\theta0}^{-1})^2)
        = 2 \sum_{ij} \Tr(V_{i\xi_i} T_i V_{j \xi_j} T_j),
\end{align}
\end{widetext}
and since $T_i = S^{-1}_{0 \xi_i}A_{\xi_i} Q_{\xi_i}$ and $V_{i\xi_i}$ are symmetric:
\begin{align}
    \dv{V_{i \xi_i}} E[\naive] &= \dv{V_{i \xi_i}} \sum_j Tr(V_{j \xi_j} T_j)
        = T_i ,\\
    \dv{V_{i \xi_i}} \Var[\naive] &= 2 \sum_{jk} \dv{V_{i \xi_i}} \Tr(V_{j \xi_j} T_j V_{k \xi_k} T_k) .
\end{align}
We apply the same algorithm as described previously,
except that we need to use the elements of $T_i$ instead of $P_\xi$ to decide which elements of $V_{i \xi_i}$ to set and which value to use.
The scaling factor is then calculated as in \autoref{eq:derate}.

\section{Application to a neutrino model tune}

\begin{figure*}[p]
    \centering
    \includegraphics[width=0.45\textwidth]{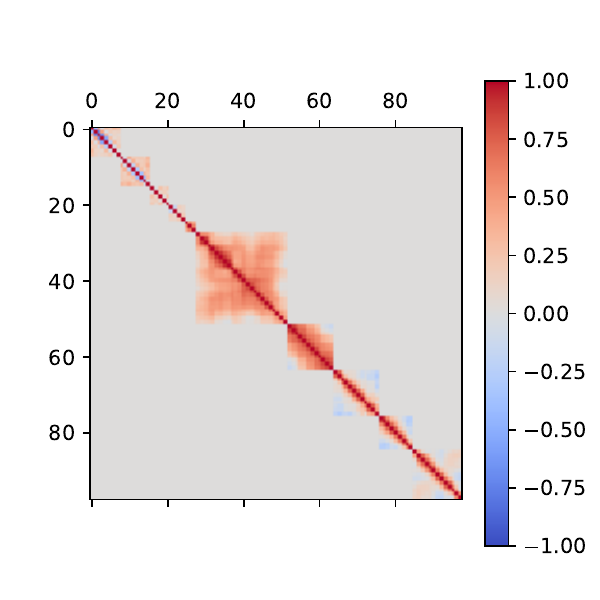}

    \includegraphics[width=0.45\textwidth]{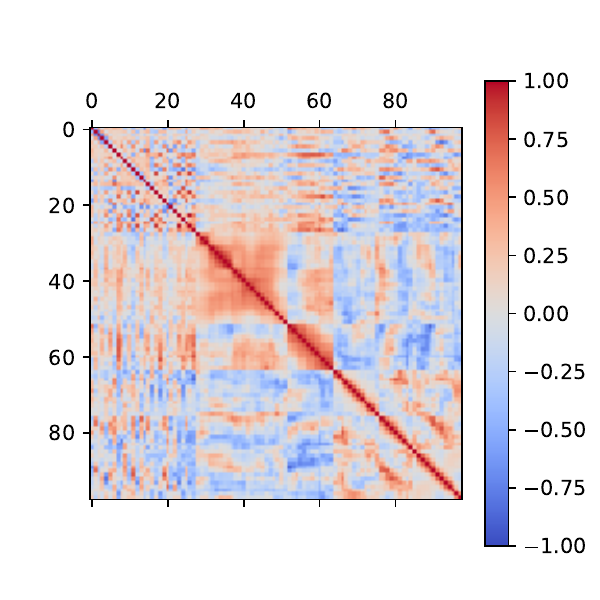}
    \includegraphics[width=0.45\textwidth]{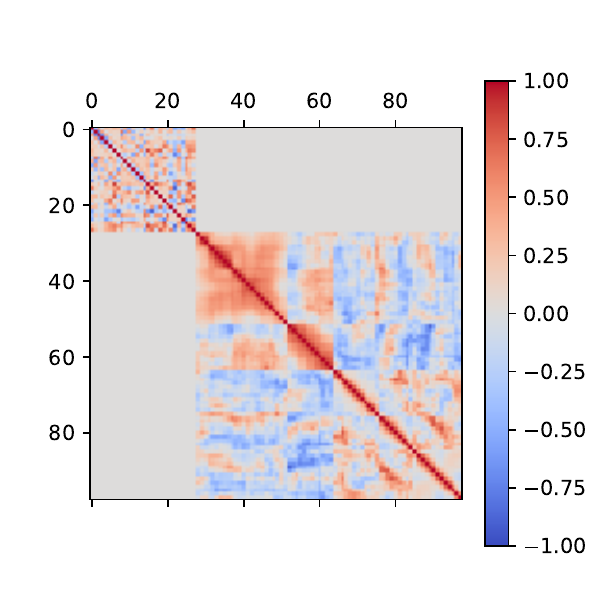}
    \caption{%
        Data correlation matrices before the norm-shape-transform for the ``RedPar'' fit done in \cite{Li2024} (top), provided by \cite{Li}.
        Applying the algorithm described in the main text yields the nightmare correlations (bottom left), and derating factors for the resulting parameter covariance of $3.87$.
        When assuming no correlations between the results from T2K and MINERvA (bottom right),
        the resulting derating factors is $2.70$.
    }
    \label{fig:genie-cor}
\end{figure*}

In \cite{Li2024}, the authors fit different combinations of model parameters in the neutrino event generator GENIE\cite{AlvarezRuso2021} against multiple measurements from T2K\cite{Abe2018,Abe2021} and MINERvA\cite{Lu2018,Cai2019,Coplowe2020}.
They decompose the single results' into a norm and shape part, to get around Peelle's Pertinent Puzzle\cite{Smith2007}, but they do not address potential correlations between the results. They treat them as uncorrelated, as can be seen in the used correlation matrices in \autoref{fig:genie-cor}.

The Covariance matrix and Jacobian matrix at the best fit point was provided by \cite{Li}.
When the nightmare derating algorithm is applied for a critical confidence level of $99.7\%$ and without further assumptions, it yields a derating factor of $3.87$ for the ``RedPar'' fit.
It could be argued that the results of a single experiment are very reasonably strongly correlated, while the results of different experiments -- in this case T2K and MINERvA -- are not correlated.
Under this assumption, the derating factor is $2.70$.
The corresponding correlation matrices are shown at the bottom of \autoref{fig:genie-cor}.

This means that, in order to ensure conservative coverage up to the critical confidence level, the parameter uncertainties have to be inflated by a factor between $\sqrt{2.70} = 1.64$ and $\sqrt{3.87} = 1.97$, depending on the fit and the assumptions about the inter-experiment correlations.
This is a significant increase in uncertainty, but the best fit point is of course not affected by this, by construction of this method.

\section{Goodness of Fit and composite hypothesis tests}
\label{sec:GoF}

The covariance derating method described above allows us to define a conservative confidence interval for model parameters.
But when doing parameter fits, we often also would like to make statements about the general "Goodness of Fit" of the model.
E.g., when assuming Gaussian uncertainties and that we know the covariance perfectly well, we would calculate the squared M-distance between the data and the best fit point
\begin{align}
    GOF &= (\bm x - \bm{\hat x})^T S^{-1}_0 (\bm x - \bm{\hat x}),
\end{align}
and compare it with the expected $\chi^2_{N-k}$ distribution.

The relation to the $\chi^2_{N-k}$ distribution is more obvious when expressing the $GOF$ value in terms of the projection matrix from before:
\begin{widetext}
\begin{align}
    GOF &= (\bm x - (P (\bm x - \bm x_0) + \bm x_0))^T S^{-1}_0 (\bm x - (P (\bm x - \bm x_0) + \bm x_0)) \nonumber \\
        &= (\bm x - \bm x_0)^T (I - P)^T S^{-1}_0 (I - P) (\bm x - \bm x_0) \nonumber \\
        &= (\bm x - \bm x_0)^T R^T B^T S^{-1}_0 B R (\bm x - \bm x_0) \nonumber \\
        &= (\bm\phi - \bm\phi_0)^T S^{-1}_{\phi 0} (\bm\phi - \bm\phi_0). \label{eq:nuisance}
\end{align}
\end{widetext}
Here $(I - P)$ is the ``residual maker'' matrix, another projection matrix with an ($N-k$)-dimensional image space, which corresponds to the kernel, or null space, of the original projection matrix.
Since it is a projection matrix, it can be expressed as $(I - P) = BR$, with an $N \times (N-k)$ matrix $B$ and $R = (B^T S^{-1}_0 B)^{-1}B^T S^{-1}_0$,
as long as the columns of $B$ span the null space of $P$.
$R$ thus projects into the ``null parameter space'' $\bm\phi = R \bm x$, the details of which do not matter.
The ``null parameters'' are an arbitrary set of $(N-k)$ parameters that explore the data space in $\bm x$ that is not reachable by the $k$ model parameters $\bm\theta$.
The columns of $B$ can be constructed from $A$, for example by doing a Singular Value Decomposition.

The structure of \autoref{eq:nuisance} is identical to \autoref{eq:parameters};
we just replaced the model parameter Jacobian $A$ with the ``null parameter'' Jacobian $B$.
We can thus use the exact same method as for the parameter estimation, to determine a derating factor for the Goodness of Fit calculation.
This also means that we can use the derating method for composite hypothesis ests.
In that case, we simply derate the Goodness of Fit test for a fit of all free parameters of the composite hypothesis as described above.

\section{Conclusions}

In this paper we discussed several ways to deal with data sets with missing information about the correlation between some or all of the data points.
The general class of $\fmax$ statistics is a potential alternative to the $\naive$ M-distance when doing simple hypothesis tests.
The $\fitted$ and $\pmin$ test statistics are examples of this.
The latter is especially easy to apply to the combination of multiple experiments.
If the smallest p-value among the experiments is small, the combined p-value is simply that smallest p-value multiplied by the number of combined experiments.
The $\fmaxopt$ statistic tries to ensure a good statistical power by minimizing the maximum total naive M-distance that is still accepted for any given confidence level.

When doing parameter fits, the $\fmax$ statistics are not very suitable, as they are not smoothly differentiable everywhere, and the expected distributions of the best fit point values are unknown.
For this purpose it is possible to derate the naive M-distance to ensure a conservative coverage of the true values even if unknown correlations are present in the data.
This does not affect the best-fit value of the parameters, only the resulting uncertainties.
The value of the derating factor depends on the structure of the known covariance and the data subspace that is covered by the model parameters.
An algorithm to estimate the worst-case factor to achieve robustness against correlations up to any given confidence level is available\cite{Koch2024}.
Applied to the test case of the fits done in \cite{Li2024}, it shows that the parameter uncertainties must be inflated by up to factor $1.97$, depending on the assumptions about the correlations.
This is a significant increase in uncertainty, and shows how important it is to provide -- and use -- proper correlations between results, if they are to be used together.

The same method can be applied to do a conservative Goodness of Fit or composite hypothesis test.
Instead of the projection matrix, which projects any possible data point to the fitted model expectation, it uses the ``residual maker'' matrix, which projects into the null space of the original projection; it projects out the residual of the fit.
With this, the method produces a derating factor to make the Goodness of Fit or composite hypothesis test conservative up to a chosen confidence level, just like in the parameter estimation case.
In general, the necessary derating factor for the parameter estimation and the Goodness of Fit test are \emph{not} identical.

In no particular order:
I would like to thank all participants of the NuXTract 2023 workshop at CERN for the discussions that sparked the ideas that led to this paper.
I would like to thank Stephen Dolan, and Callum Wilkinson for providing feedback which made this paper more useful than it would be otherwise.
Special thanks go to Weijun Li for patiently explaining the details of the GENIE tune to me and providing me with the additional inputs needed to apply my method, Marco Roda for pointing me in the right direction to use the Professor framework, and Julia Tena Vidal for helping me understand their norm-shape-transform.
This work was funded by the Deutsche Forschungsgemeinschaft (DFG, German Research Foundation) under Germany’s Excellence Strategy -- EXC 2118 PRISMA+ -- 390831469.

\bibliography{biblio}%

\end{document}